\documentclass[conference,hidelinks]{IEEEtran}
\IEEEoverridecommandlockouts
\usepackage{cite}
\usepackage{amsmath,amssymb,amsfonts}
\usepackage{todonotes}
\usepackage{algorithmic}
\usepackage{graphicx}
\usepackage{textcomp}
\usepackage{xcolor}
\definecolor{airforceblue}{rgb}{0.36, 0.54, 0.66}
\definecolor{brinkpink}{rgb}{0.98, 0.38, 0.5}
\def\BibTeX{{\rm B\kern-.05em{\sc i\kern-.025em b}\kern-.08em
    T\kern-.1667em\lower.7ex\hbox{E}\kern-.125emX}}
    
\usepackage{listings}
\usepackage{hyperref}
\usepackage{enumitem}

\ifCLASSOPTIONcompsoc
\usepackage[caption=false,font=normalsize,labelfont=sf,textfont=sf]{subfig}
\else
\usepackage[caption=false,font=footnotesize]{subfig}
\fi
    
\usepackage{aas_macros}

\newcommand{\orcid}[1]{\href{https://orcid.org/#1}{\includegraphics[height=10pt]{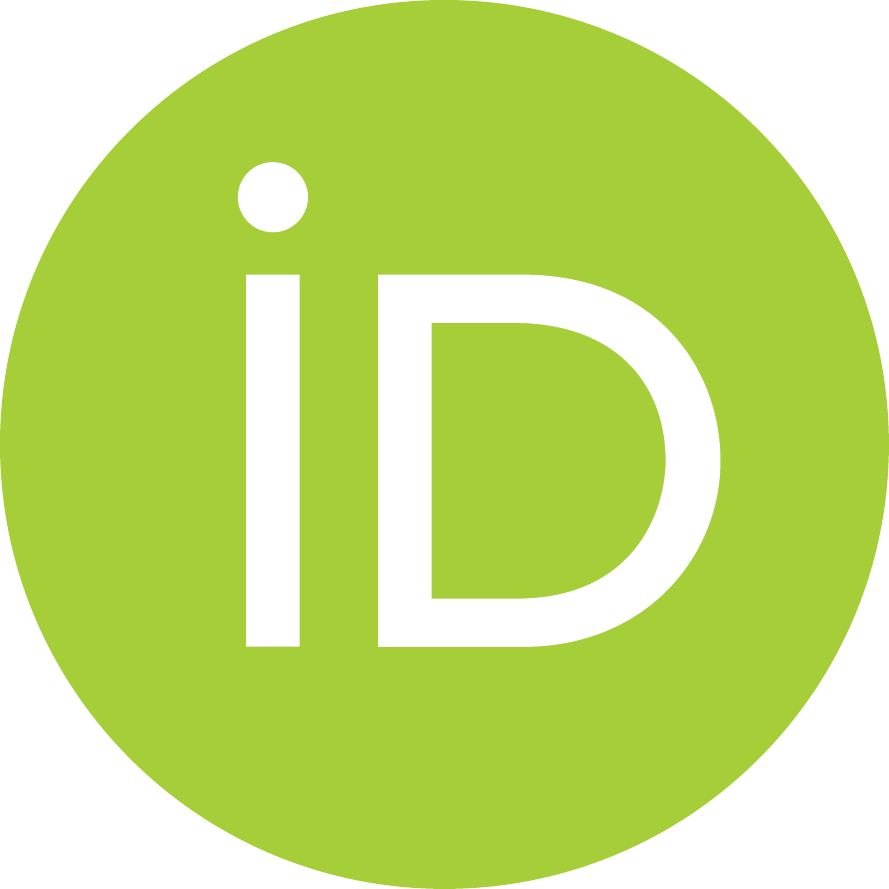}}}

\begin{document}
\bstctlcite{IEEEexample:BSTcontrol}

\title{From Merging Frameworks to Merging Stars: Experiences using HPX, Kokkos and SIMD Types}

\author{
Gregor Dai\ss\orcid{0000-0002-0989-5985}\IEEEauthorrefmark{2}, Srinivas Yadav Singanaboina\IEEEauthorrefmark{1},
\IEEEauthorblockN{Patrick Diehl\orcid{0000-0003-3922-8419}\IEEEauthorrefmark{1}\IEEEauthorrefmark{4}, Hartmut Kaiser\orcid{0000-0002-8712-2806}\IEEEauthorrefmark{1}, Dirk Pfl\"uger\orcid{0000-0002-4360-0212}\IEEEauthorrefmark{2}}
\IEEEauthorblockA{
\IEEEauthorrefmark{1}LSU Center for Computation \& Technology, Louisiana State University,
Baton Rouge, LA, 70803 U.S.A}
\IEEEauthorblockA{\IEEEauthorrefmark{2} IPVS, University of Stuttgart,
Stuttgart, 70174 Stuttgart, Germany\\
Email: Gregor.Daiss@ipvs.uni-stuttgart.de}
\IEEEauthorblockA{\IEEEauthorrefmark{4} Department of Physics and Astronomy, Louisiana State University,
Baton Rouge, LA, 70803 U.S.A. }
}

\maketitle

%Octo-Tiger, a large-scale 3D AMR code for the merger of stars, uses a combination of HPX, Kokkos and explicit SIMD types, aiming to achieve performance-portability for a broad range of heterogeneous hardware.
%However, on A64FX CPUs, we encountered several missing pieces, hindering performance by causing problems with the SIMD vectorization.
%Therefore, we add std::experimental::simd as an option to use in Octo-Tiger's Kokkos kernels alongside Kokkos SIMD, and further add a new SVE backend.
%Additionally, we amend missing SIMD implementations in the Kokkos kernels within Octo-Tiger's hydro solver.
%We test our changes by running Octo-Tiger on three different CPUs: An A64FX, an Intel Icelake and an AMD EPYC CPU, evaluating SIMD speedup and node-level performance.
%Using our changes, we get a good SIMD speedup on the A64FX CPU, as well as noticeable speedups on the other two CPU platforms. However, we also experience a scaling issue on the EPYC CPU.

\begin{abstract}
Octo-Tiger, a large-scale 3D AMR code for the merger of stars, uses a combination of HPX, Kokkos and explicit SIMD types, aiming to achieve performance-portability for a broad range of heterogeneous hardware.
However, on A64FX CPUs, we encountered several missing pieces, hindering performance by causing problems with the SIMD vectorization.
Therefore, we add std::experimental::simd as an option to use in Octo-Tiger's Kokkos kernels alongside Kokkos SIMD, and further add a new SVE (Scalable Vector Extensions) SIMD backend.
Additionally, we amend missing SIMD implementations in the Kokkos kernels within Octo-Tiger's hydro solver.
We test our changes by running Octo-Tiger on three different CPUs: An A64FX, an Intel Icelake and an AMD EPYC CPU, evaluating SIMD speedup and node-level performance.
We get a good SIMD speedup on the A64FX CPU, as well as noticeable speedups on the other two CPU platforms. However, we also experience a scaling issue on the EPYC CPU.
\end{abstract}

%\begin{IEEEkeywords}
%HPX, Kokkos, SIMD, SVE, AVX512, parallel programming models, high-level abstractions, astrophysical simulations
%\end{IEEEkeywords}

\section{Introduction}
Astrophysical applications drive the need for high-performance computing.
They require a lot of computational power for their simulation, as well as numerous developer hours to ascertain they run efficiently on the ever-changing set of current hardware platforms.

Octo-Tiger is an astrophysical simulation used to simulate binary star systems and their eventual outcomes~\cite{marcello2021octo}. %\todo[inline]{Patrick: Do we have some recent astro results to cite here? No, they did not publish any paper since that paper.}
Octo-Tiger is built on HPX~\cite{kaiser2020hpx} for task-based programming to scale to all cores within one compute node and, beyond that, to thousands of other compute nodes in distributed scenarios.
Using this, Octo-Tiger achieved scalability on Cori~\cite{heller2019harnessing}, and more recently Piz Daint~\cite{daiss2019piz_short} and Summit~\cite{diehl2021octo}.

Consequently, given these last two machines, the most recent work on Octo-Tiger was focused on porting the original CPU-only implementation of Octo-Tiger to GPUs.
To have portable GPU kernels, much of this was done with Kokkos and an HPX-Kokkos integration to allow us to use Kokkos kernels as HPX tasks. 
To improve CPU performance of said Kokkos kernels, we used SIMD (Single Instruction Multiple Data) types provided by Kokkos~\cite{sahasrabudhe2019portable}. While working well in Octo-Tiger, this was only used in the Kokkos kernels of one solver and only tested on older CPU platforms~\cite{daiss2021beyond}.
%To improve CPU performance of the Kokkos kernels in Octo-Tiger's gravity solver, explicit vectorization is used, utilizing the SIMD types provided by Kokkos~\cite{sahasrabudhe2019portable, daiss2021beyond}.

In turn, this work focuses not on running Octo-Tiger on GPUs, but instead on our development process modifying Octo-Tiger to run efficiently on Fujitsu A64FX\textsuperscript{\texttrademark} CPUs. 
% Factcheck: That's how fujitsu refers to their on cpu (with trademark): 
% https://www.fujitsu.com/global/about/resources/news/press-releases/2018/0822-02.html
% Should hopefully be correct...
This is done to prepare Octo-Tiger for experimental runs on Fugaku.%, but for the development work we are using a test allocation on Stony Brook University's A64FX Ookami system.

%To improve CPU performance of the Kokkos kernels in Octo-Tiger's gravity solver, we previousely had good experiences using explicit vectorization with the Kokkos SIMD types

Hence, we will use the current development snapshot of Octo-Tiger to investigate how well our utilized execution model (a mixture of HPX, Kokkos, and explicit vectorization) works on modern CPUs and share some additions we had to make when porting to A64FX.
Although said mixture of frameworks already gained us a certain degree of portability, we had to supplement some missing ingredients in this work to make Octo-Tiger run more efficiently on A64FX CPUs and make use of their SVE SIMD instructions:
%\begin{enumerate}
\begin{enumerate}[topsep=0pt,itemsep=-1ex,partopsep=1ex,parsep=1ex]
    \item To have a wider range of SIMD backends accessible, we integrated \lstinline{std::experimental::simd} within our Kokkos kernels (while maintaining compatibility with the Kokkos SIMD types).
    \item We add \lstinline{std::experimental::simd} compatible SVE types, allowing for explicit vectorization on A64FX CPUs.
    \item Some of our Kokkos kernels simply did not yet use the explicit vectorization with SIMD types, as they have only been used on GPUs so far. Therefore, we supplement those kernels (two kernels in our new hydro solver) with these types and apply SIMD masking where required.
\end{enumerate}
To test these new additions, we run node-level tests on an A64FX CPU, determining node-level scaling and SIMD speedup using various backends (both for the entire application and the most important CPU kernels). We investigate the speedup of both the new hydro SIMD implementation and the speedup of the surrounding hydro solver in a hydro-only compute scenario.

We further use this opportunity to run the same tests (using different SIMD types) on recent Intel and AMD CPUs, to evaluate the performance on current CPUs, as previous CPU results were gathered on now-outdated hardware with fewer cores~\cite{daiss2021beyond}. Moreover, those previous results did not yet include the SIMD additions to the hydro solver from this work.

This paper is structured as follows: 
Section~\ref{sec:application} briefly introduces the real-world astrophysics application: Octo-Tiger. 
Section~\ref{sec:technology} emphasizes the technology used in Octo-Tiger and how its components fit together to create a powerful yet portable execution model. Section~\ref{sec:changes} describes the changes that we had to make for this to work for A64FX. 
Section~\ref{sec:results} lists the performance measurements conducted with the real-world application. 
Section~\ref{sec:related:work} focuses on the related work. 
Finally, Section~\ref{sec:conclusion} concludes the work and outlines the next steps.

\section{Scientific application: Octo-Tiger}
\label{sec:application}
From the astrophysics perspective, our interest are binary star systems and their eventual outcomes, especially white dwarf mergers and the contact binary v1309 Sco. For the latter, emission of the red light during the merger was observed. To gain more understanding in the process, self-consistent simulations with a very high resolution to resolve the star atmosphere to extract the simulated light curve are necessary. This would allow for a direct comparison with the observation data. Furthermore, with more understanding of the light curve in v1309 Sco, it will be possible to reliably simulate the light curve of other star mergers.

In the following, we show a brief overview of the solvers and data-structure Octo-Tiger uses to simulate these binary star systems.

\begin{figure}[tb]
    \centering
    \includegraphics[width=0.8\linewidth]{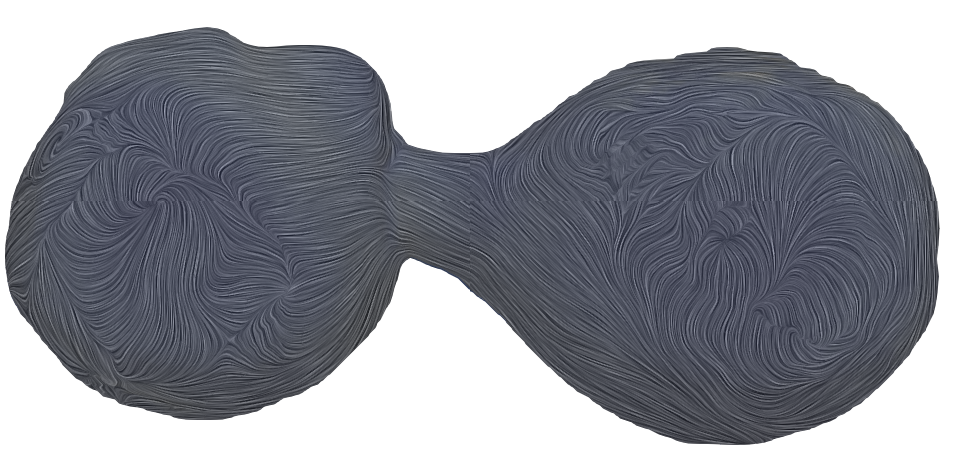}
    \caption{Flow on the surface of the two stars. In the center we see the aggregation belt with the mass transfer from the smaller star to the larger star.}
    \label{fig:octo_example}
\end{figure}

\paragraph{Solvers}
The star system is modeled as a self-gravitating, astrophysical fluid. To solve the system, Octo-Tiger uses a coupled hydro solver and gravity solver. The inviscid Navier-Stokes equations are solved using finite volumes in the hydro solver~\cite{kurganov2000}. In turn, Newtonian gravity is solved using a modified Fast Multipole Method (FMM) in the gravity solver~\cite{octotiger_fmm}. 

The most important Octo-Tiger compute kernels are part of those two solvers. In the hydro solver, there is the \texttt{reconstruct} kernel, using the piecewise-parabolic method computing the values of the variables that are being evolved at 26 quadrature points. Furthermore, there is the \texttt{flux} kernel, which uses those values to compute the final flux using Newton-Cotes quadrature.
In turn, the gravity solver contains multiple compute kernels calculating the same-level interactions between close-by cells (the most compute-intensive FMM step): The \texttt{Monopole} kernel for non-refined cells and the \texttt{Multipole} kernel (and its rho variants for angular momentum correction and root variant for a specialization for the root sub-grid).
For brevity, we refer to~\cite{marcello2021octo} for more details on the solvers. For more details on gravity solver kernels in particular, we refer to~\cite{Pfander18acceleratingFull,daiss2018octo}. Lastly, for a convergence study of a production run and more details about the hydro kernels, we refer to~\cite{diehl2021performance}.

\paragraph{Data-Structure}
Octo-Tiger uses adaptive mesh refinement (AMR) to focus on the area of interest, the atmosphere between the two stars with the aggregation belt, see Figure~\ref{fig:octo_example}. For efficient computations, an adaptive octree with a Cartesian sub-grid in each leaf is used. Each sub-grid has $512$ cells within a $8\times 8 \times 8$ cube, which is the default. Hence, each compute kernel invocation is operating on one sub-grid only, using HPX to launch many such kernels concurrently for the available sub-grids.
The cube size of the sub-grids can be configured during compile time. The performance benefits of various sub-grid sizes were studied in~\cite{diehl2021octo}.

\section{HPX, Kokkos and SIMD Types}
\label{sec:technology}
Octo-Tiger is built using a combination of frameworks that make up our execution model:
We use HPX for task-based parallelization and distributed computing.
We use Kokkos for writing compute kernels that are portable between various CPU and GPU platforms.
Lastly, we use SIMD types (provided by Kokkos) for explicit vectorization on CPUs, whilst still supporting GPU execution (via instantiation of the SIMD template type with scalar types).

In this section, we briefly cover these utilized frameworks and how they can fit together to complement each other.

\subsection{HPX}

%\begin{itemize}
HPX is an Asynchronous Many-Task Runtime system that is implemented in C\texttt{++}~\cite{kaiser2020hpx,kaiser_hartmut_2022_6969649}. 
The library implements all APIs related to concurrency and parallelism as mandated by recent ISO standards C\texttt{++}20 and C\texttt{++}23 in a conforming way. 
In this context, it implements all the (more than $100$) parallel algorithms as described in the C\texttt{++} standard.
%
%It has been described in detail in other publications, such as~\cite{hpx_pgas_2014,Kaiser:2015:HPL:2832241.2832244,Heller2016}. 
It has been described in detail in other publications, such as~\cite{Kaiser:2015:HPL:2832241.2832244,Heller2016}. 
%
%In the context of this paper, HPX has been used for two purposes:
%
%a) to provide a framework allowing to extend the existing parallel algorithms to support explicit vectorization, and \todo{we do not use the parallel algorithms here}
%
%b) as an underlying runtime platform for the Octo-Tiger astro-physics application used as the domain science driver (see~\cite{10.1145/3295500.3356221} for more details), thus managing local and distributed parallelism while observing all necessary data dependencies. 
%Data and task dependencies can be expressed with HPX futures, and chained together an execution graph. This graph can be build asynchronously, with the HPX worker threads processing the tasks when there dependencies are fullfilled. This task-based programming model is particulary useful for parallel implementations of adaptive, tree-based codes like Octo-Tiger, as we can build the task graph quickly to make concurrent work available to the system.

In the context of this paper, HPX has been used as an underlying runtime platform for the Octo-Tiger astrophysics application used as the domain science driver (see~\cite{daiss2019piz_short} for more details), thus managing local and distributed parallelism while observing all necessary data dependencies. 
Data and task dependencies can be expressed with HPX futures, and chained together in an execution graph. This graph can be built asynchronously, with the HPX worker threads processing the tasks when their dependencies are fulfilled. This task-based programming model is particularly useful for parallel implementations of adaptive, tree-based codes like Octo-Tiger, as we can build the task graph quickly when traversing the tree to make concurrent work available to the system.

We also use the performance monitoring library APEX~\cite{10.1145/2491661.2481434} that is well integrated with HPX. 
APEX can be applied to capture a combination of task-based events with hardware counter information for optimizing HPX on different hardware platforms. 
APEX can further measure the runtime of annotated HPX tasks, getting mean execution times for them, which can be used to determine speedups for specific parts of the code (such as compute kernels with and without SIMD).
%\item AMT
%\item Mention APEX
%\end{itemize}
\subsection{HPX with Kokkos}
For portability, we use Kokkos'~\cite{9485033} abstractions for memory and execution within Octo-Tiger. 
Kokkos allows us to easily write compute kernels that work on both CPU and GPU, allowing us to run the same kernel implementation on NVIDIA\textsuperscript{\textregistered} and AMD\textsuperscript{\textregistered} GPUs, and even on the CPU if desired.

Usually, these kernels would be launched in a fork-join manner using Kokkos, for example using an OpenMP execution space on the CPU to achieve concurrency.
However, there exist two HPX-Kokkos integrations, which allow us to avoid this and use Kokkos in a more task-based fashion. 

The first one is the Kokkos HPX execution space. Kernels launched within this execution space are split into HPX tasks, which will be processed by the existing HPX worker threads (hence there is no need for multiple competing thread pools).
However, this alone does not provide the functionality to launch Kokkos functions asynchronously and integrate them within HPX's execution graph. 
For that, we use the second integration: HPX-Kokkos~\cite{daiss2021beyond}. This integration (and its executors) allow us to obtain \lstinline{hpx::futures} for asynchronous Kokkos kernel (and function) launches, facilitating their integration with HPX's asynchronous execution graph.

% This part is very verbose - delete?
Together, these integrations allow us to write portable Kokkos kernels that we can launch from arbitrary HPX tasks and treat as HPX tasks themselves. 
This way, we can express dependencies between kernels (and other tasks) using HPX futures, automatically triggering new tasks when an asynchronously launched kernel finishes. Furthermore, this way the Kokkos kernels make use of HPX resources for CPU execution (namely the existing worker threads).
%Together, this not only allows us to write portable kernels using Kokkos, but also to launch them from arbitrary HPX tasks, to treat them themselves as HPX tasks (allowing us to express dependencies via HPX futures, automatically triggering tasks that use the results of a kernel) and lastly, to make them use the HPX resources for CPU execution (namely the existing worker threads).

\subsection{Explicit Vectorization with the Kokkos SIMD types}
To gain truly portable kernels that run well on both the CPU and GPU, we need to take SIMD vectorization into account.
%Modern CPUs offer a lot of the potential floating-point performance by being able to run instructions on multiple data items at once (Single Instruction Multiple Data, or SIMD).
Modern CPUs offer a lot of the potential floating-point performance because of their ability to run instructions on multiple data items at once (SIMD).
%\todo[inline]{Introduce SIMD abbreviation in related work?}
While the compiler can use this automatically in certain circumstances, it is often more reliable to use the appropriate instructions directly to ensure SIMD usage. Since different hardware platforms often use different instruction sets (AVX, AVX512, NEON and SVE to name some examples), abstractions using C\texttt{++} types have been developed to increase portability and ease-of-use. Kokkos itself includes such SIMD types\footnote{https://github.com/kokkos/simd-math}, notably capable of using vectorization on the CPU by compiling to the correct SIMD instructions, but also able to use scalar types in case the same compute kernel is compiled for GPU usage~\cite{sahasrabudhe2019portable}.

\subsection{Utilizing the Frameworks within Octo-Tiger}

\begin{figure}[t]
  \centering
  \includegraphics[width=.999\linewidth]{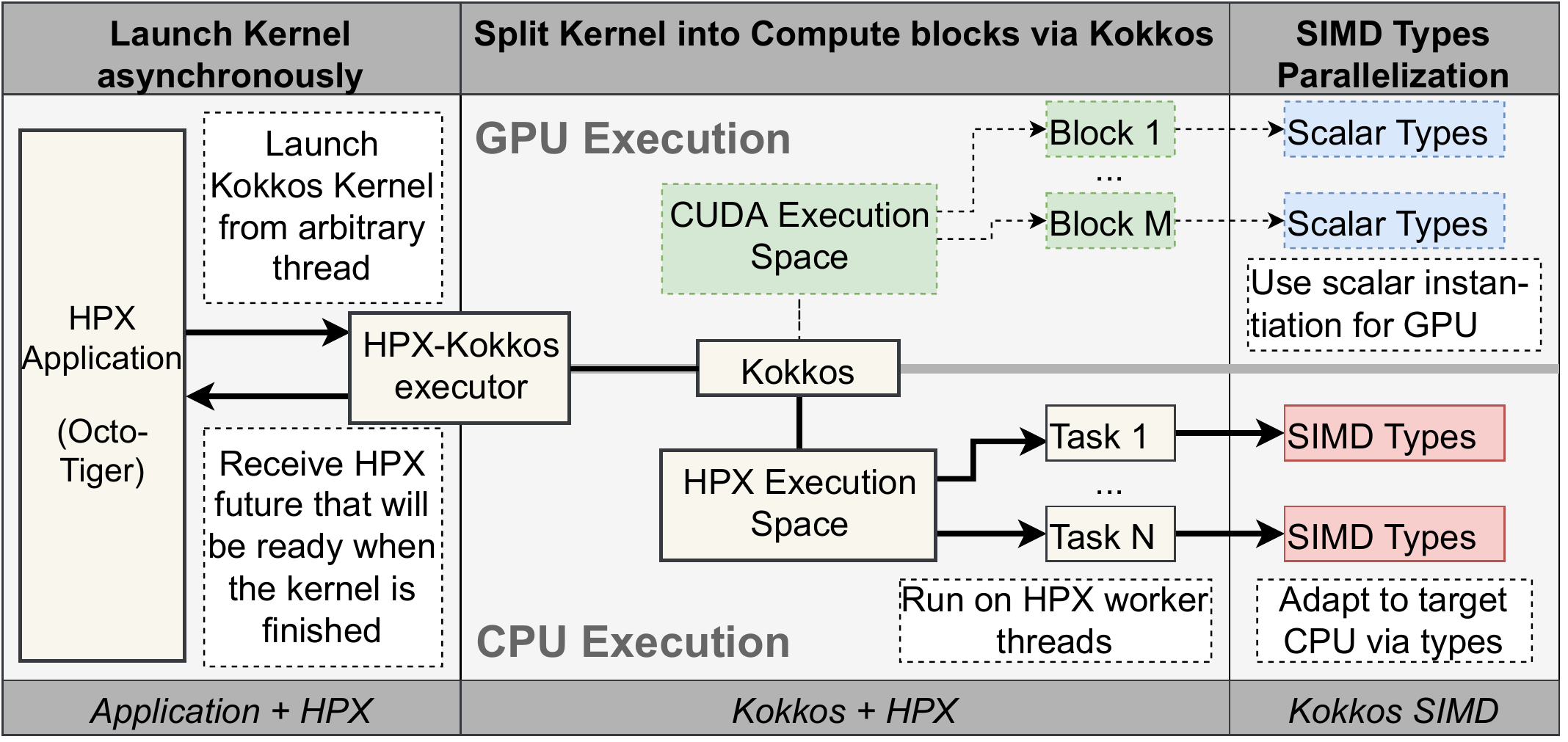}  
  %\caption{Schematic of the way Octo-Tiger launches compute kernels: We launch Kokkos kernel concurrently from arbitrary HPX tasks (and thus threads), keeping track of the results with the HPX futures HPX-Kokkos returns. Kokkos in turn splits the kernel into HPX tasks (using the HPX execution space). Inside those tasks we use SIMD types for explicit SIMD vectorization (whilst keeping the kernel compatible with GPU exeuction.}
    \caption{The execution model used to launch compute kernels in Octo-Tiger: We launch Kokkos kernels concurrently from arbitrary HPX tasks (and thus threads), keeping track of the results via the HPX futures HPX-Kokkos returns. Kokkos in turn splits the kernel into HPX tasks (using the HPX execution space). We use SIMD types for explicit SIMD vectorization whilst keeping the kernel compatible with GPU execution.}
    \label{fig:execution_model}
\end{figure}

Octo-Tiger was built to rely on HPX for parallelization and distributed computing, with each sub-grid in the octree representing an HPX component which can be placed on arbitrary compute nodes. However, its compute kernels changed significantly over the last two years.
In the past, Octo-Tiger used separate compute kernels for CPU execution (using Vc for SIMD types) and for the GPU execution (using NVIDIA\textsuperscript{\textregistered}~CUDA\textsuperscript{\textregistered})~\cite{daiss2019piz_short,daiss2018octo}.
%In an effort to unify these implementations into a single portable one for each kernel, we started to slowly switch to Kokkos at the same time (and in the same work) that saw the introduction of the HPX-Kokkos integrations~\cite{daiss2021beyond}.
In an effort to unify these implementations into a single portable one for each kernel, we started to slowly switch to Kokkos when we developed and introduced the general idea and framework of the HPX-Kokkos integrations~\cite{daiss2021beyond}.
 
As mentioned in Section~\ref{sec:application}, we have multiple kernels for each solver: The gravity solver contains \texttt{multipole} and \texttt{monopole} kernels (and various specializations that include angular momentum corrections (rho variant) and a specialization for the tree root subgrid to process remaining gravity interactions not handled by sub-grids down the tree. 
The hydro solver includes the \texttt{reconstruct} and the \texttt{flux} kernels.

Each kernel only operates on one sub-grid (and its ghost-layers) at a time. This means that during the solver execution, we traverse the tree, launching a multitude of different Kokkos compute kernels in the process.
Using the HPX-Kokkos integration, these kernels can be launched (for different sub-grids) from multiple tasks simultaneously, with their execution status and results being integrated into the HPX execution graph using the futures returned by these asynchronous launches.

Depending on which device the Kokkos kernels use for execution, it will internally use either scalar types (on GPU) or types compiling down to the appropriate SIMD instructions (for example AVX512 for Intel) -- provided the kernel has been implemented with the SIMD types. 
Of course, we can control which SIMD types are being used at compile time, meaning we can easily use scalar types on the CPU as well if we want to gauge the speedup we gain by using SIMD.

This model of an HPX application launching and synchronizing compute kernels using HPX / HPX-Kokkos, executing them via Kokkos on the correct device in the correct block configuration and, finally, having them adapted to the target device with SIMD types is exemplified in Figure~\ref{fig:execution_model}.

\section{Software Changes and Additions for this work}
\label{sec:changes}
Preparing Octo-Tiger to run on Fugaku, we contributed multiple changes recently to improve performance on CPU platforms generally, but on A64FX CPUs particularly.
\subsection{Add SIMD types to Kokkos hydro kernels}
\label{sec:changes:kernel}
At the start of this work, the hydro compute kernels (\texttt{reconstruct} and \texttt{flux}) within Octo-Tiger were using Kokkos, but did not yet contain SIMD types, as the amount of branching in these kernels proved to be problematic for the SIMD implementation in the past.
We added a first implementation with the SIMD types to Octo-Tiger\footnote{Part of https://github.com/STEllAR-GROUP/octotiger/pull/426}, reducing branching where possible and using SIMD masks where not possible. This implementation was later refined\footnote{https://github.com/STEllAR-GROUP/octotiger/pull/428} and is currently still in the process of being further optimized.

To improve the execution time of small compute kernels (very much like the two hydro kernels) launched with the Kokkos HPX execution space, we added an optimization that if a kernel is small enough to be launched as one task, it will be executed immediately on the same core (as to not waste the hot cache). We hope to upstream this change to Kokkos soon.

\subsection{Integrate \lstinline{std::experimental::simd} SIMD types}
%\begin{itemize}
%    \item What is it?
%    \item Short description what we needed to change to add it to Octo-Tiger (i.e. how easy it actually was)
%    \item Shortcoming of the approach (Opens up for using std::simd backends, but we are still limited to the Kokkos SIMD API as we have to stay compatible to %both  for device-side compilation)
%\end{itemize}
\lstinline{std::experimental::simd} is an explicit vectorization API which was proposed for target Parallelism TS 2 for C\texttt{++}20 standard.
\lstinline{std::experimental::simd} is inspired by the Vc library (now deprecated) and provides support for explicit vectorization using data-parallel types.
These are high-level C\texttt{++} abstract types which are inlined to vector intrinsics of the underlying CPU architecture. 
This helps in writing portable high-level vectorized code using data-parallel types (and associated helper functions), without having to deal with intrinsics and low-level code, whilst still obtaining the same performance benefits from the vectorization.

Octo-Tiger supports vectorization using Kokkos SIMD and thus adding a \lstinline{std::experimental::simd} vectorization backend was straightforward as it was just a change of the namespace from Kokkos SIMD to \lstinline{std::experimental::simd}. Additionally, we only needed one wrapper for the \texttt{choose} function due to a difference in the method name in between the two frameworks. 
With this small integration\footnote{https://github.com/STEllAR-GROUP/octotiger/pull/403}, we can essentially still use exactly the same Kokkos compute kernels, but instantiate them with the SIMD types from \lstinline{std::experimental::simd} instead.

However, we have not completely switched to \lstinline{std::experimental::simd} because the Kokkos SIMD types also work for GPU execution (being instanced as scalar double types). This does not seem to be supported yet in \lstinline{std::experimental::simd}, as GPU device-side compilation with their scalar types fails. 
As we continue to use both frameworks, we are limited to using only the functionality of \lstinline{std::experimental::simd} that has a counterpart in Kokkos SIMD (or has a fallback implementation we provide). 

%However, integrating \lstinline{std::experimental::simd} still opens our Kokkos kernels up to a more diverse set of possible SIMD backends (which might not be implemented within the Kokkos SIMD library or provide difficulties on a specific machine).
Integrating \lstinline{std::experimental::simd} still opens our Kokkos kernels up to a more diverse set of possible SIMD backends (as we can now use all SIMD types within \lstinline{std::experimental::simd} inside the Kokkos kernels).

\subsection{Add \lstinline{std::experimental::simd}-compatible SVE types}
Whilst the Kokkos SIMD contains SVE types itself already, these seem to rely on autovectorization (unlike different SIMD instructions sets in the library that are implemented with explicit vectorization). Furthermore, this particular SVE backend did not compile for Octo-Tiger initially, and only yielded scalar performance after having been modified to compile. 

%To still benefit from the performance boost SVE offers, we implemented a  
%\lstinline{std::experimental::simd}-compatible SVE backend (not relying on vectorization by the compiler) offering SVE SIMD types to our Kokkos kernels.
To still benefit from the performance boost SVE offers, we implemented a  
\lstinline{std::experimental::simd}-compatible SVE backend (not relying on vectorization by the compiler) offering SVE SIMD types to our Octo-Tiger Kokkos kernels and to other applications.
%While we tested this with Octo-Tiger in this work, these types can also be used in other applications using \lstinline{sve::experimental::simd}.

Our new backend offers a portable high-level API which defines most of the functionalities of \lstinline{std::experimental::simd} with the help of the ARM C Language Extensions (ACLE) for SVE defined in \texttt{arm\_sve.h}. 
This is a standard header provided by ARM\textsuperscript{\textregistered} that contains all the types and functions for SVE vectorization currently supported by GCC and Clang. 
For benefiting from SVE vectorization, we used explicit vectorization with the help of fixed length vector registers defined by the compiler. 
The vector length is set during compile time using the flag \texttt{-msve-vector-bits=N}. 
%This new SVE backend is available on GitHub\footnote{https://github.com/srinivasyadav18/sve}.

Integrating this SVE backend within Octo-Tiger is straightforward given our \lstinline{std::experimental::simd} integration: One only needs to include the header and change the SIMD namespace accordingly. Then the Kokkos kernels can use our new SVE types.
Using our experimental implementation of the SVE types enables Octo-Tiger to use SVE in all Kokkos compute kernels.

We made our new SVE backend available on GitHub\footnote{https://github.com/srinivasyadav18/sve}. Notably, it can be used independent of Octo-Tiger in other applications. These applications simply need to use its SIMD SVE types to achieve explicit vectorization on A64FX CPUs.
\section{Results}
\label{sec:results}
We use three different CPU platforms for our tests: 
\begin{itemize}[topsep=0pt,itemsep=-1ex,partopsep=1ex,parsep=1ex]
    \item A two-socket AMD\textsuperscript{\textregistered} EPYC\textsuperscript{\texttrademark} 7H12 CPU @ 2.60GHz with $128$ cores.
    \item A two-socket Intel\textsuperscript{\textregistered}~Xeon\textsuperscript{\textregistered} Platinum 8358 CPU @ 2.60GHz with $64$ cores (Intel\textsuperscript{\textregistered} Icelake node).
    \item A Fujitsu A64FX\textsuperscript{\texttrademark} CPU @ 1.80 GHz on Stony Brook University's Ookami cluster with $48$ cores (ARM\textsuperscript{\textregistered} node).
\end{itemize}
% Fact checking: Ookami is using 1.8 ghz - see https://cnc-workshop.github.io/cnc2021/slides/CnC_Siegmann.pdf
It is worth mentioning that the AMD and Intel nodes can and will run at higher frequencies as their boost is enabled.
We will run three different tests on each CPU.
For all tests, we use double precision.

For brevity, we use SES to stand for \lstinline{std::experimental::simd} in the following. For example, with SES AVX512 the Kokkos kernels use the AVX512 types within \lstinline{std::experimental::simd}, with KOKKOS NEON they use the NEON types within the Kokkos SIMD library instead.  
SES SVE makes them use our new SVE types.% introduced in the last section.

\begin{figure}[t]
\centering
\subfloat[\label{fig:simd-speedup-icelake}On the Intel Icelake node] {
\centering
  \includegraphics[width=.238\textwidth]{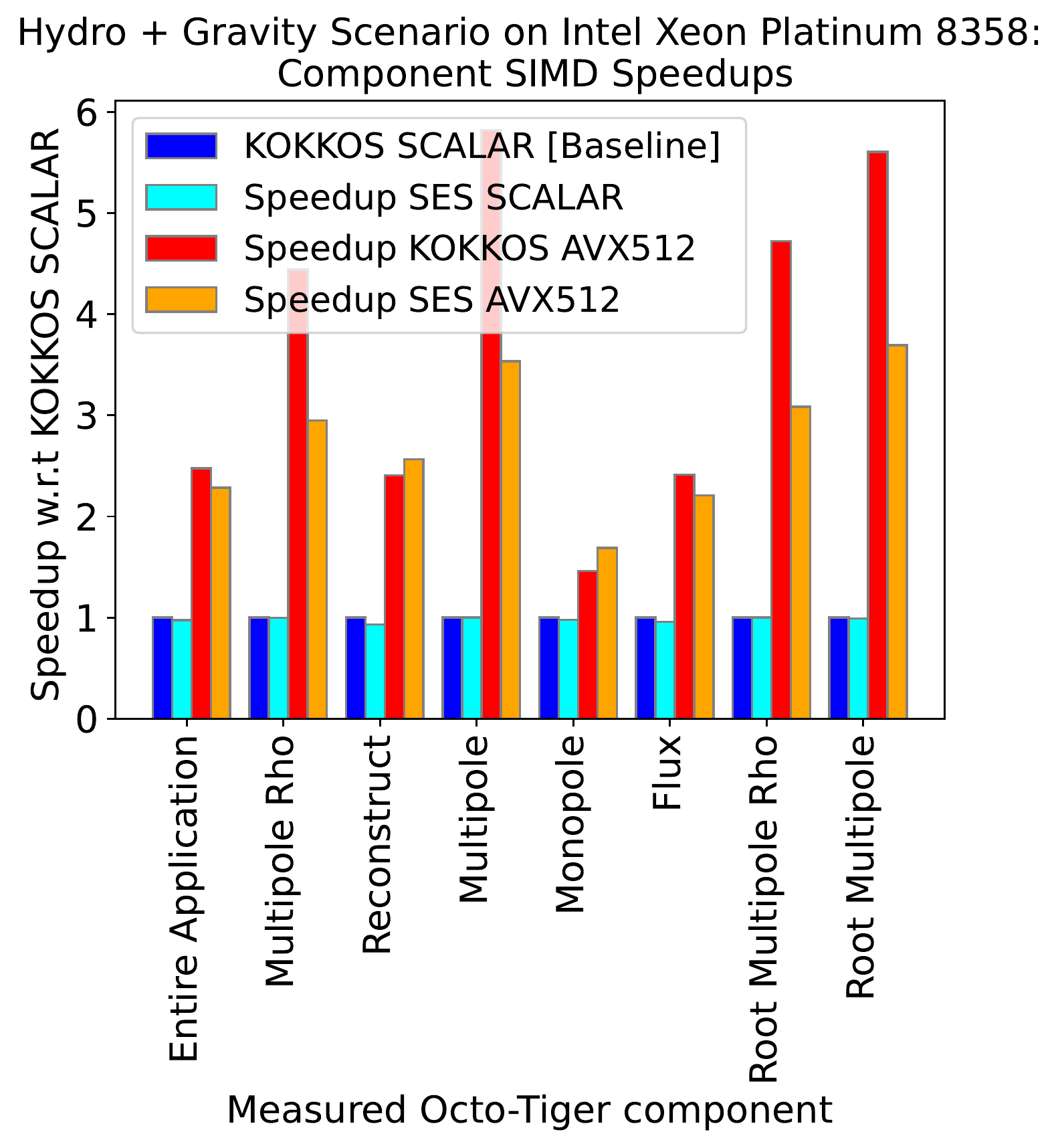}  
}
\hspace*{-0.2cm} 
\subfloat[\label{fig:simd-speedup-epyc}On the AMD EPYC node] {
\centering
  \includegraphics[width=.236\textwidth]{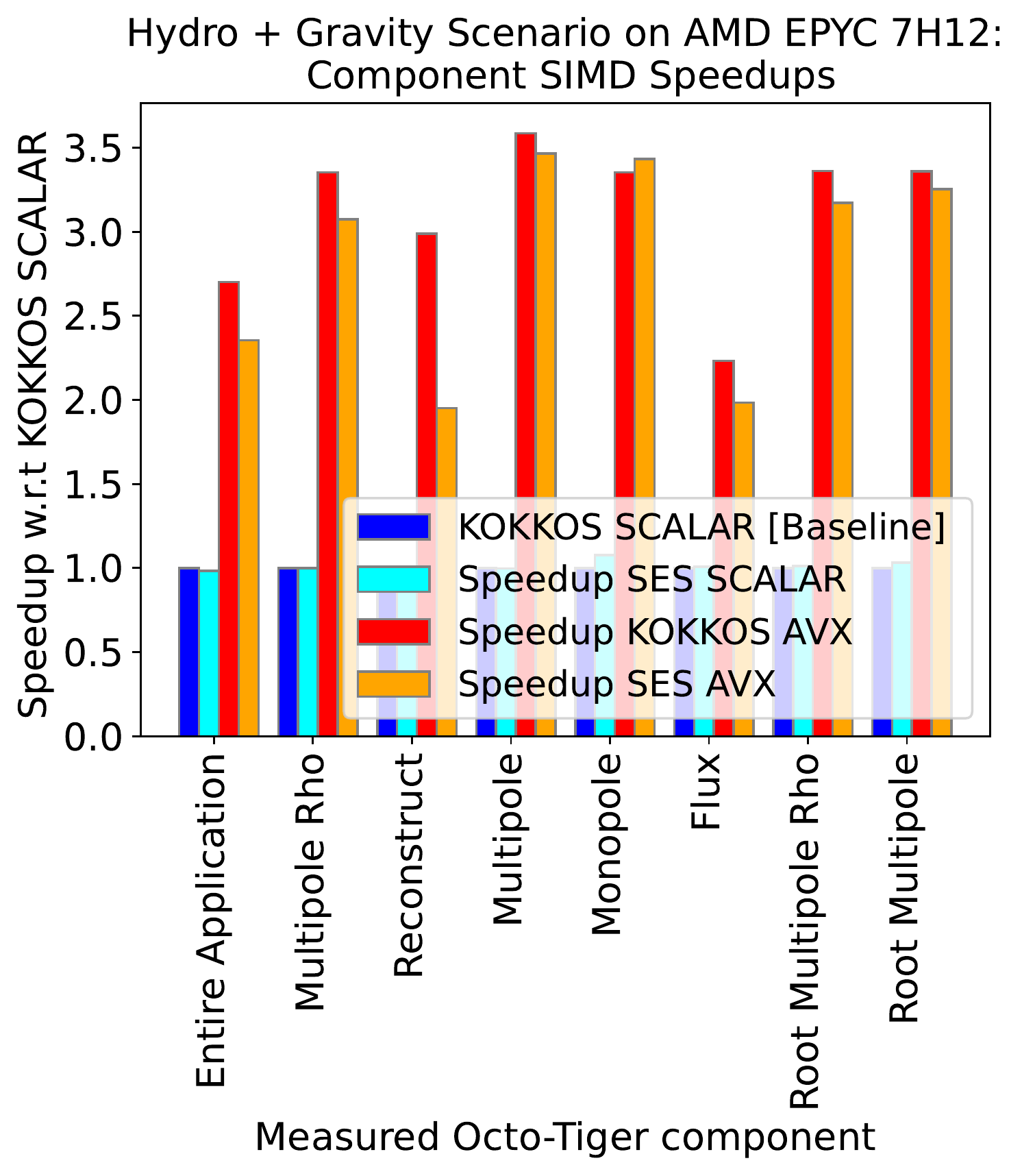}  
}

\subfloat[\label{fig:simd-speedup-neon}On A64FX (using NEON)] {
\centering
  \includegraphics[width=.238\textwidth]{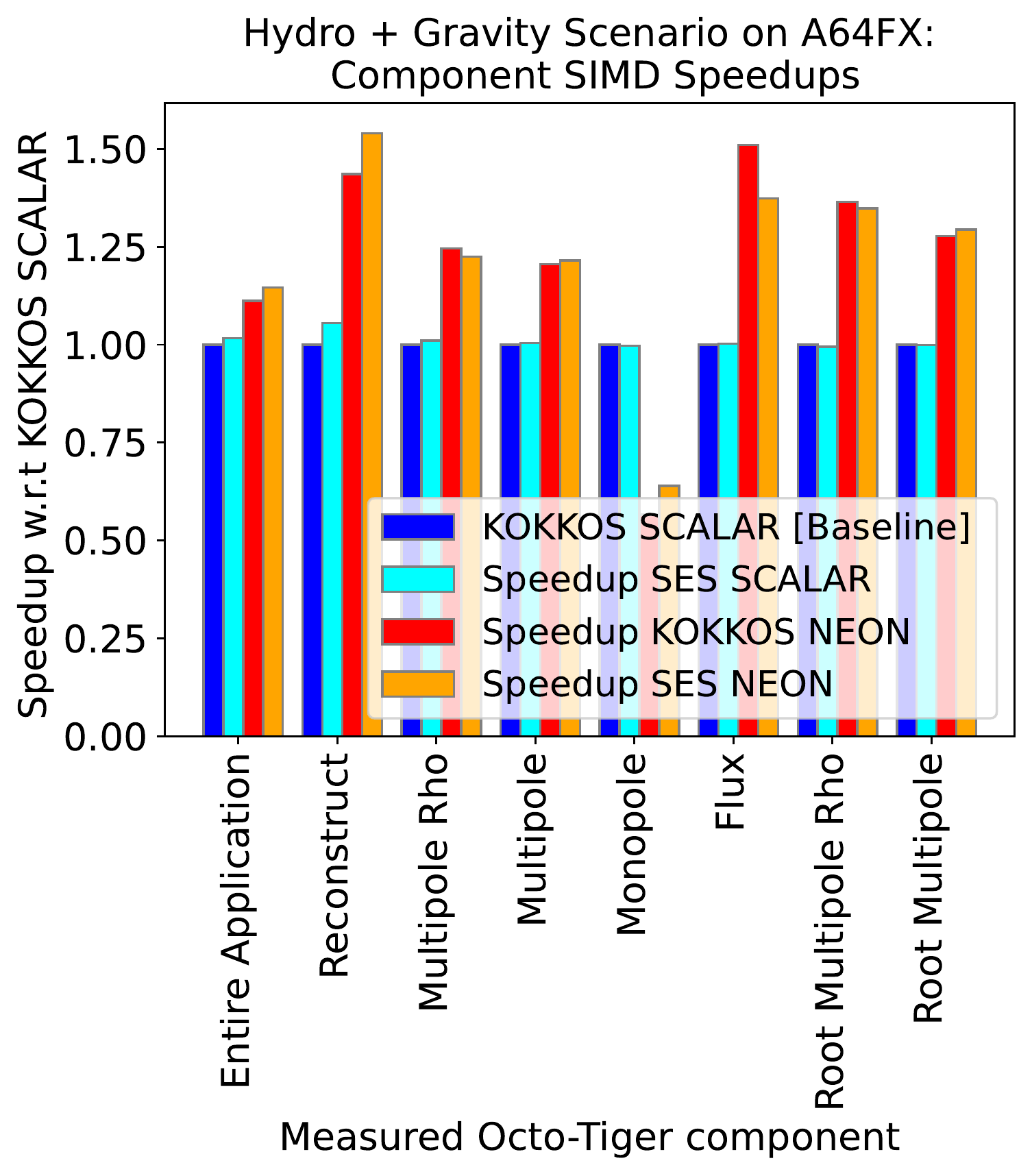}  
}  
\hspace*{-0.2cm} 
  \subfloat[\label{fig:simd-speedup-sve}On A64FX (using SVE)] {
\centering
  \includegraphics[width=.236\textwidth]{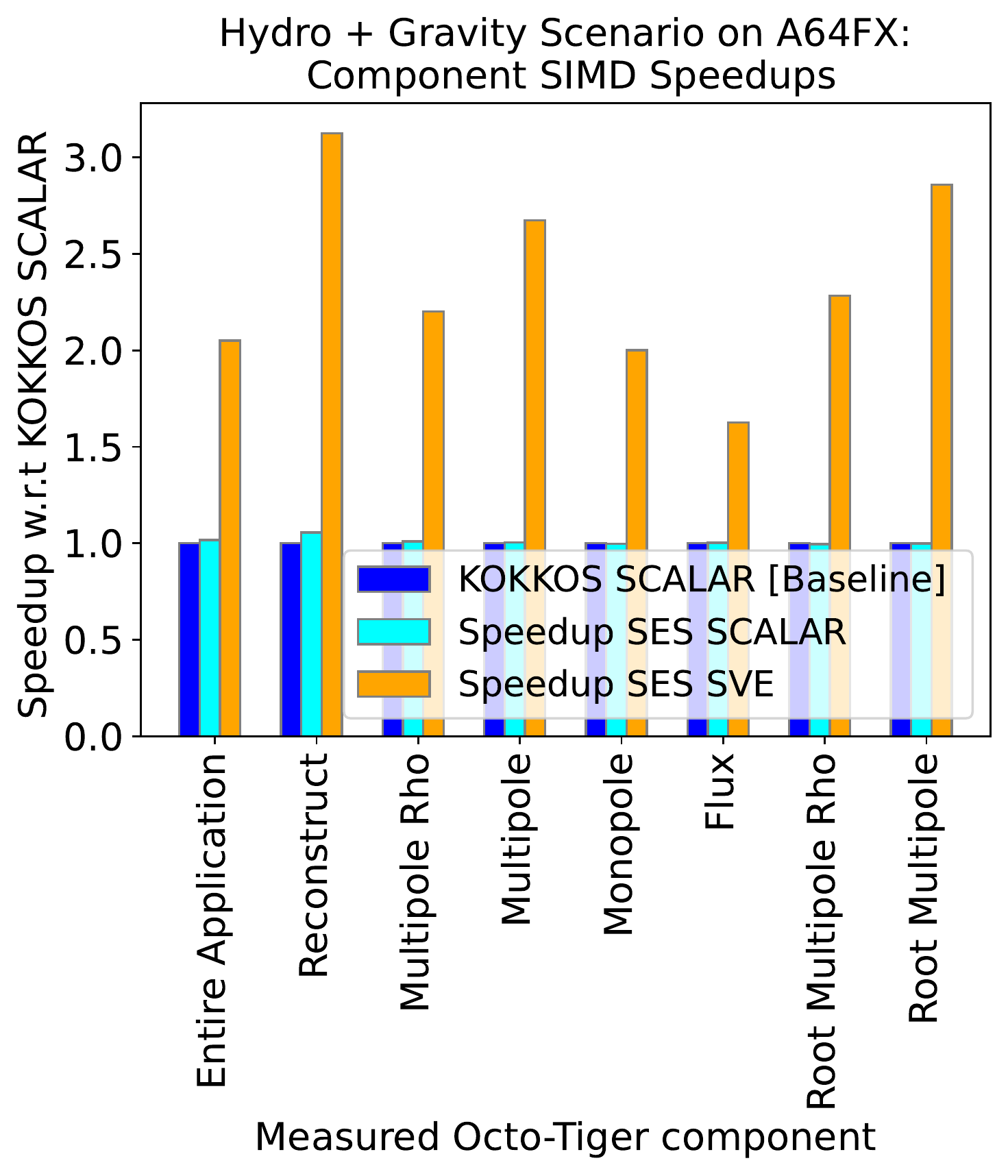}  
}  
\caption{Single Core SIMD Speedups with respect to the runtime with scalar types. The speedups are shown for both the entire application and for the relevant compute kernels.}
\label{fig:simd-speedup}
\end{figure}
\subsection{Test 1: Single-Core SIMD Speedups}
First, we take a look at the single-core speedups using SIMD.
For this, we use the "rotating star" scenario~\cite{marcello2021octo}, simulating a rotating star in equilibrium.
%For this, we use the "rotating star" scenario\footnote{Octo-Tiger parameters: \texttt{--config\_file=rotating\_star.ini} \texttt{--unigrid=1} \texttt{--max\_level=3} \texttt{--stop\_step=10} \texttt{--correct\_am\_hydro=0} \texttt{--theta=0.34} \texttt{--disable\_output=1} }, simulating a rotating star in equilibrium.
This is one of Octo-Tiger's basic tests, as it uses both the gravity and the hydro solver, helping to uncover even small errors if the simulation runs long enough.
In the context of this work, it is a useful benchmark, as it includes both solvers, and the additional work we would encounter in a production scenario (tree-traversals, exchange of ghost cells, determination of time-step size, and so on).
We disable the adaptive refinement here, to make the runtimes more comparable to a later run of a hydro-only scenario: Without AMR, both runs build a full tree for a given tree depth, resulting in the same amount of overall cells. 
We choose a low tree level to only have $512$ leaf sub-grids, as this is realistic for large distributed runs, where we also only have a limited number of sub-grids available per compute node.
%\todo[inline]{insert parameter footnote}

Given this scenario, we would like to know how both the total application speedup, as well the speedups for all major compute kernels when using SIMD.
By using APEX~\cite{10.1145/2491661.2481434}, we get the mean runtime for each compute kernel at the end of the run. 
Therefore, we simply need to compile Octo-Tiger with the SIMD (or scalar) types we want to use and run the simulation to obtain the runtime data. Over successive runs with different types, we can easily get the speedups for both the entire application and the compute kernels themselves this way.
The baseline for our speedups will be the runtime using the Kokkos scalar types (KOKKOS SCALAR). 
Note that we do not disable the compiler autovectorization, as we want a fair estimate of the benefit of our explicit vectorization using SIMD types compared to what the compiler is able to autovectorize.

Besides these baseline runtimes, we run with the \lstinline{std::experimental::simd} scalar types (SES SCALAR). They should give no speedup, but can be useful to uncover unexpected bottlenecks.
Finally, we run with the appropriate SIMD types on each CPU, using both \lstinline{std::experimental::simd} and the Kokkos SIMD types, calculating the speedup using the previous runtimes of the Kokkos SCALAR runs.

Figure~\ref{fig:simd-speedup} shows the speedups for each compute kernel and the entire application runtime when switching to the SIMD types. 
Each set of bars in a subfigure represents on Octo-Tiger component, with the bars themselves representing the speedup using one of the possible types.
There are four subfigures, one for the Intel CPU (Fig.~\ref{fig:simd-speedup-icelake}), one for the AMD CPU (Fig.~\ref{fig:simd-speedup-epyc}), and two for the A64FX CPU.
Here, on A64FX, we test the SIMD implementation both with NEON (Fig.~\ref{fig:simd-speedup-neon}) and with our new experimental SVE types (Fig.~\ref{fig:simd-speedup-sve}). 
While there are also SVE types within the Kokkos SIMD library, we were unable to make them work beyond scalar performance. Hence, these Kokkos SVE types are missing in Figure~\ref{fig:simd-speedup-sve}, and we only use the experimental SES SVE types introduced in this work.

Overall, we get a noticeable SIMD speedup for our real-world application.
While not yet as good as the more compute-intensive gravity compute kernels (i.e. \texttt{multipole} and others), the two new hydro SIMD kernels \texttt{flux} and \texttt{reconstruct} still provide noticeable speedups, especially on the A64FX machine.

On one core of the Intel Icelake, the Kokkos SIMD types (KOKKOS AVX512) perform consistently better than \lstinline{std::experimental::simd} (SES AVX512), which exemplifies that differences in the backend implementation can lead to significantly different performance outcomes. 
Using the Kokkos SIMD types for AVX and AVX512 is generally preferable, at least for our use-case.
However, the \lstinline{std::experimental::simd} interface and the SVE types we have added (SES SVE) are essential on the A64FX machine, giving us a good overall application speedup as we can now use explicit SVE vectorization.

%\todo[inline]{Rethink/rephrase the following paragraph?}
Unfortunately, the speedups we obtained fell short of perfect SIMD speedups.
However, given that our compute kernels are many hundred lines of code each, and this is a work in progress, the results are still promising (and useful for our production-runs as they already reduce the amount of compute time).
Furthermore, the flexibility of being able to simply switch between SIMD backends using types is already extremely useful here. 
Not only to be able to run on all these different platforms without changing any code to begin with, but also to switch to \lstinline{std::experimental::simd} where required (i.e.\ on the A64FX node).

\subsection{Test 2: Node-Level Scaling and Parallel Efficiency}
\begin{figure}[t]
\centering
\subfloat[\label{fig:node-level-scaling-icelake}On the Intel Icelake node] {
\centering
  \includegraphics[width=.242\textwidth]{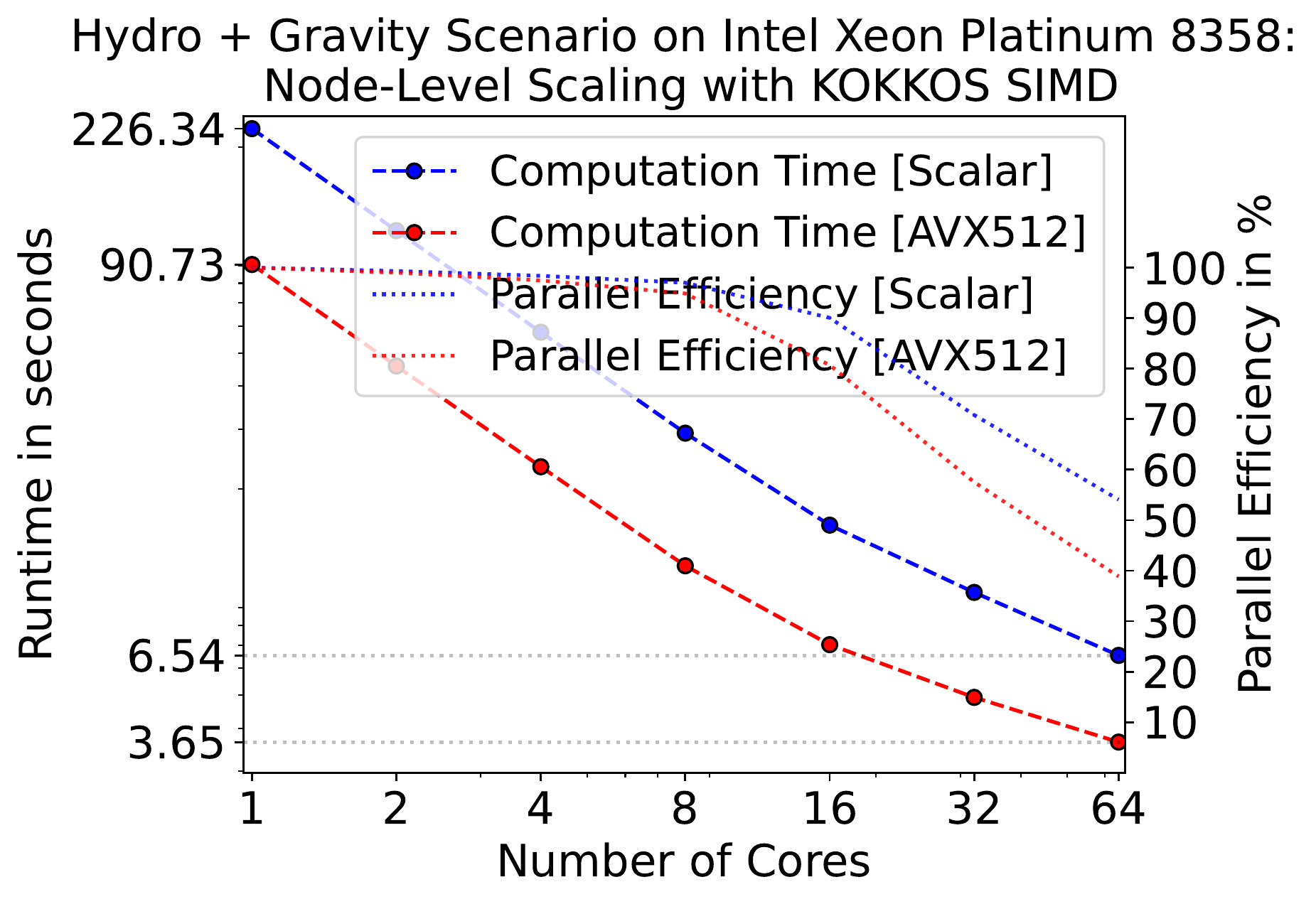}  
}
\hspace*{-0.3cm} 
\subfloat[\label{fig:node-level-scaling-epyc}On the AMD EPYC node] {
\centering
  \includegraphics[width=.242\textwidth]{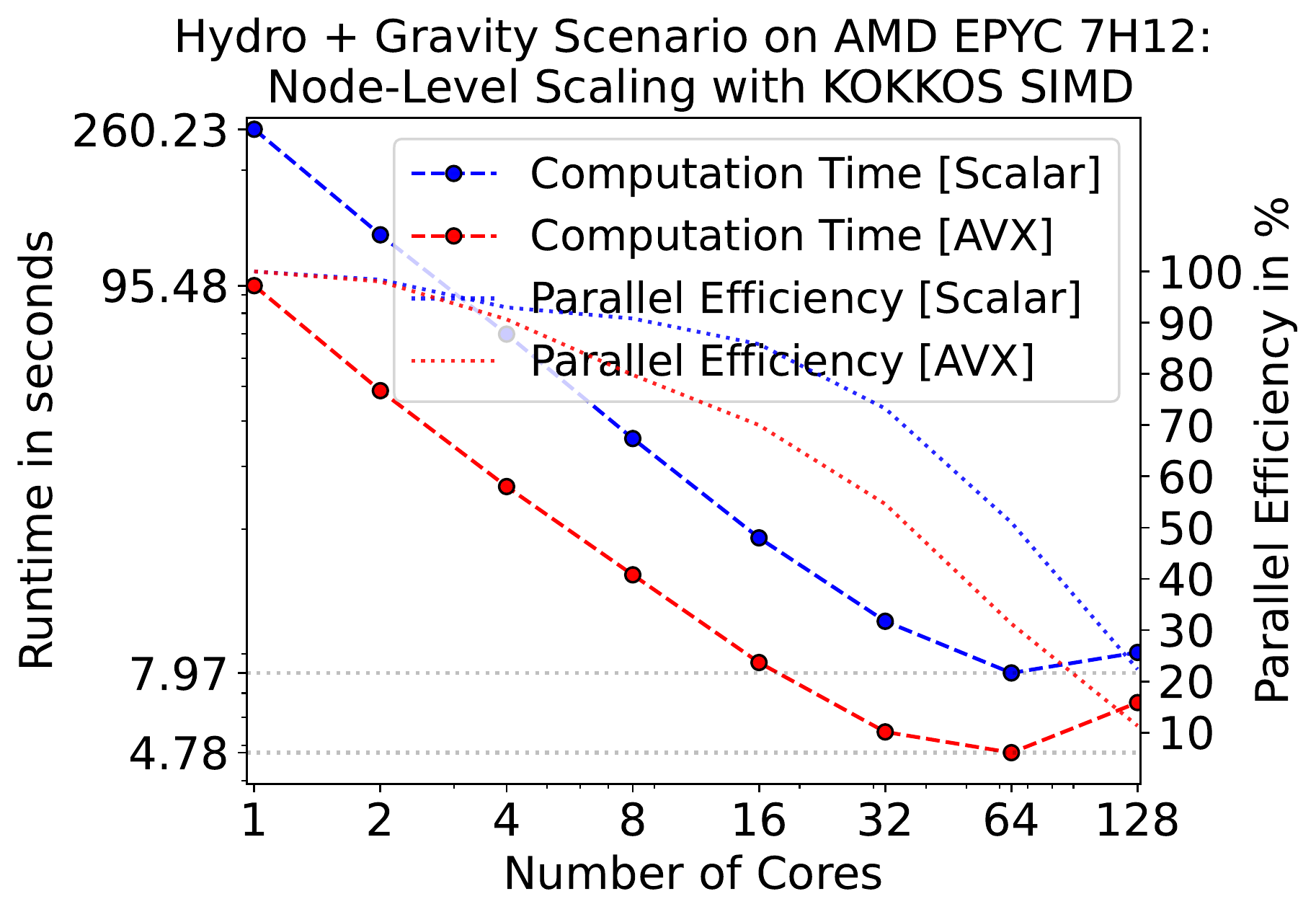}  
}

\subfloat[\label{fig:node-level-scaling-neon}On A64FX (using NEON)] {
\centering
  \includegraphics[width=.242\textwidth]{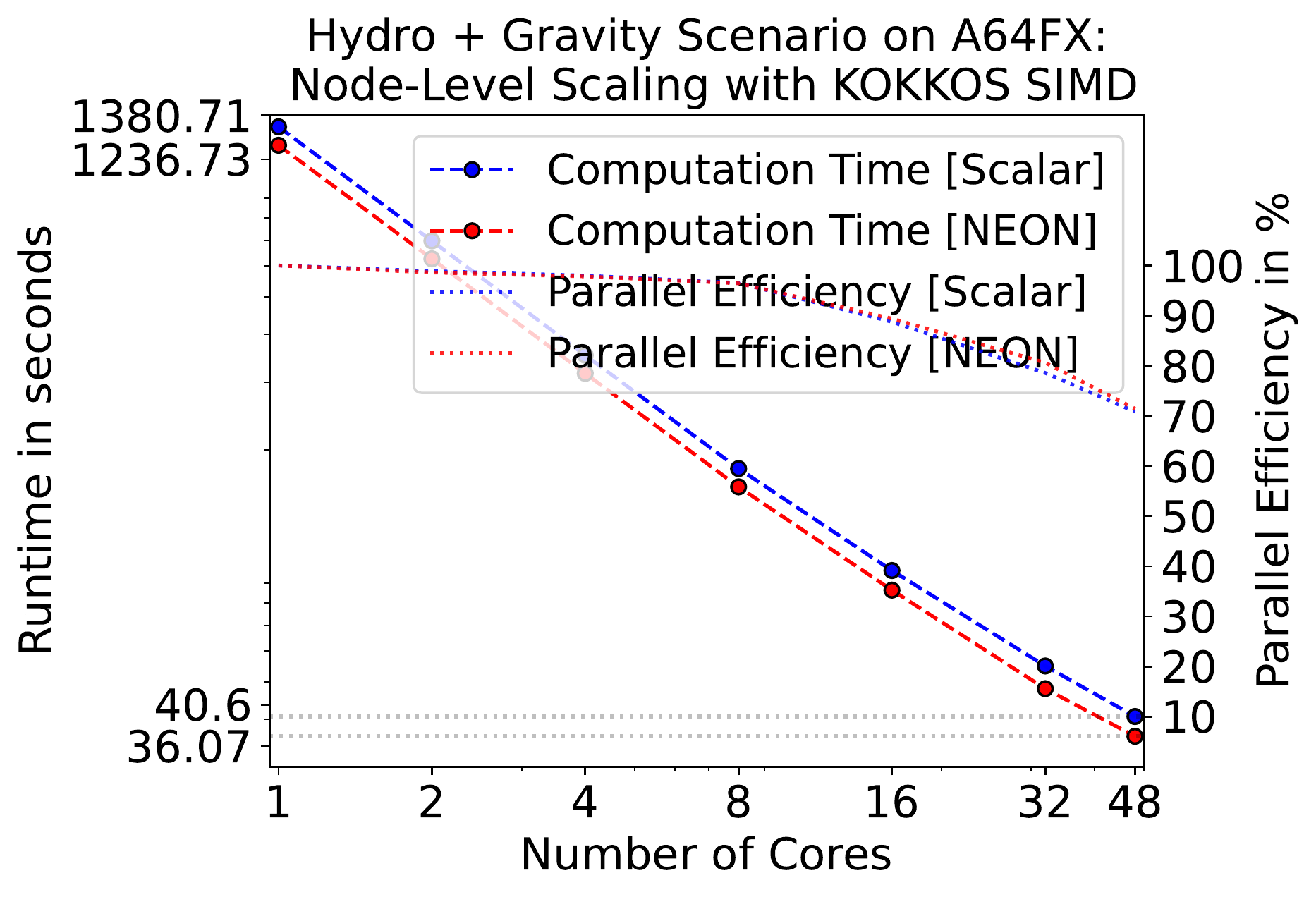}  
}
\hspace*{-0.3cm} 
\subfloat[\label{fig:node-level-scaling-sve}On A64FX (using SVE)] {
\centering
  \includegraphics[width=.242\textwidth]{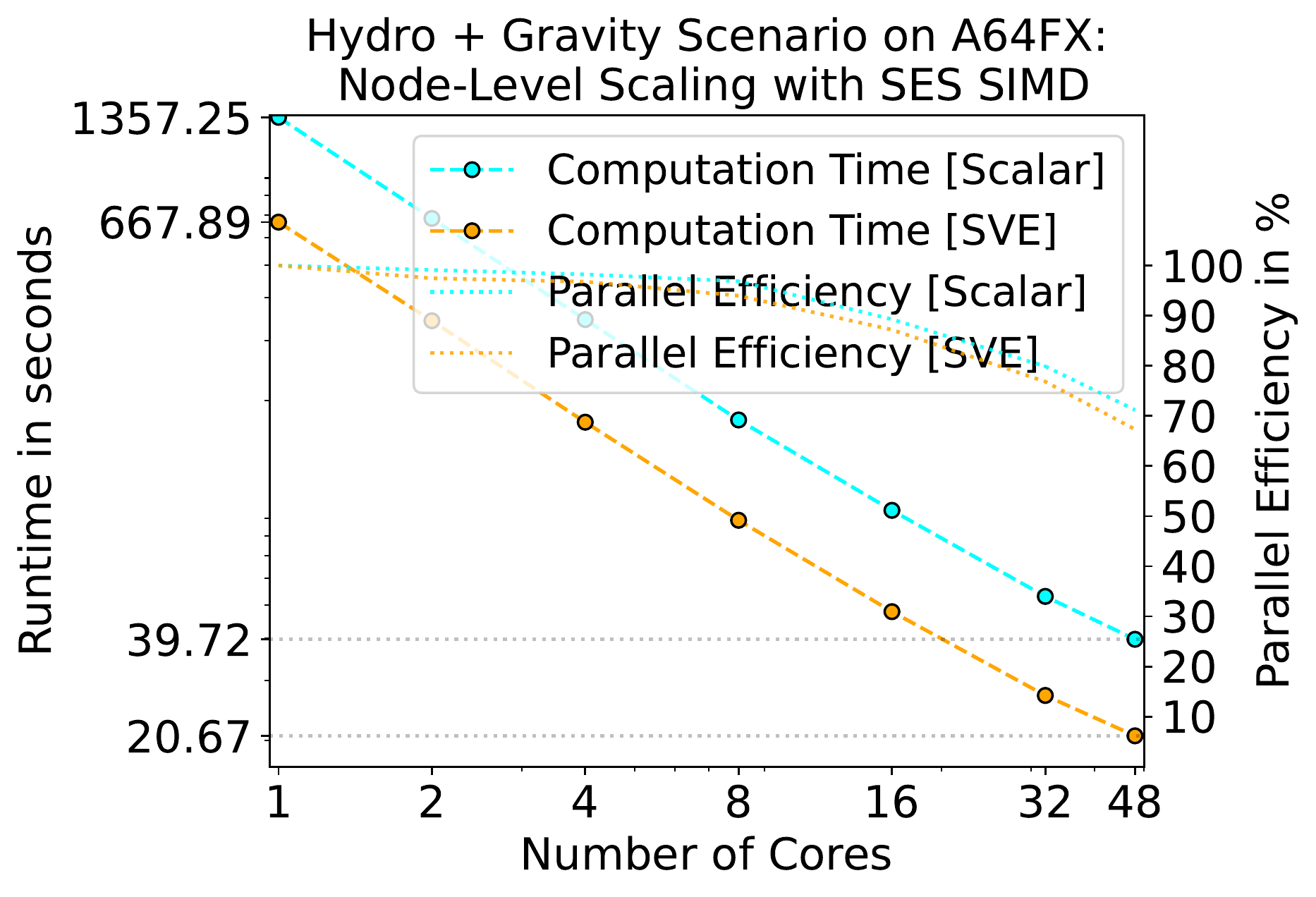}  
}
\caption{Parallel efficiency and node-level scaling, going from one to all cores of the respective three platform.
Each graph includes one run with the SIMD types for the current platform (two graphs for A64FX to show both NEON and SVE), as well the same run using the scalar types for comparison.}
\label{fig:node-level-scaling}
\end{figure}

Moving beyond one core, we now take a look at the node-level scaling.
For this, we are once again using the same Octo-Tiger scenario as for the previous test, looking at the overall computation time for 10 timesteps (each timestep consists of 6 gravity and 3 hydro solver iterations).
We are using the SIMD types with the best results from the previous test on each machine: Kokkos AVX512 on Intel, Kokkos AVX on AMD, and SES SVE on the A64FX ARM node. 
We further include a run using Kokkos NEON on this ARM node for comparison. 
For all runs, we also compare with scalar runs, to see if (and how) the SIMD speedup is influenced when scaling to the complete node: The SIMD speedup can change for a multitude of reasons (for example: frequency changes, work starvation, memory bottlenecks).

The results are mixed and can be seen in Figure~\ref{fig:node-level-scaling}.
The overall best runtime is achieved on the Intel Icelake machine (Fig.~\ref{fig:node-level-scaling-icelake}), reaching 3.65s.
Runtimes on the A64FX machine improved drastically with our new SVE backend (Fig.~\ref{fig:node-level-scaling-sve}), both over using the scalar types and over using the Kokkos NEON types (Fig.~\ref{fig:node-level-scaling-neon}).
However, the performance still falls short of the Icelake CPU. 
While this is to be expected to a certain degree given the lower processor frequency and fewer cores, the difference is noticeable. 
Looking at the APEX runtime data and comparing them to runs on different machines, it seems that every component in Octo-Tiger runs a bit slower, with no obvious new hotspot on the A64FX CPU. The slowdown is most severe in the gravity solver, though, probably due to the better SIMD speedups we get using AVX512. Furthermore, the frequency penalty on the Intel CPU when using AVX512 is lower than we expected; in fact, the CPU seems to run faster than the base frequency (staying in a range of 2.6-3.2 Ghz during test runs with our scenario).
%Looking at the average kernel runtime, we noticed that the \texttt{multipole} kernels are taking a lot longer on the A64FX machine. 
%Normally, the input of this kernel fits into the L2 cache which might be the issue on the ARM machine as we do not have a large per-core L2 cache here anymore, but have a shared L2 cache instead (which can introduce contention if other cores are streaming a lot of data for other parts of the code which happens a lot since we have to convert the input datastructure for the gravity solver from Array-of-Struct to Struct-of-Array for legacy purposes). 
%This was already an issue on the shared L2 per compute tile on the Knights Landing processors in the past~\cite{Pfander18acceleratingFull}.

\begin{figure}[t]
\centering
\subfloat[\label{fig:fixed_intel}On the Intel Icelake node] {
\centering
  \includegraphics[width=.242\textwidth]{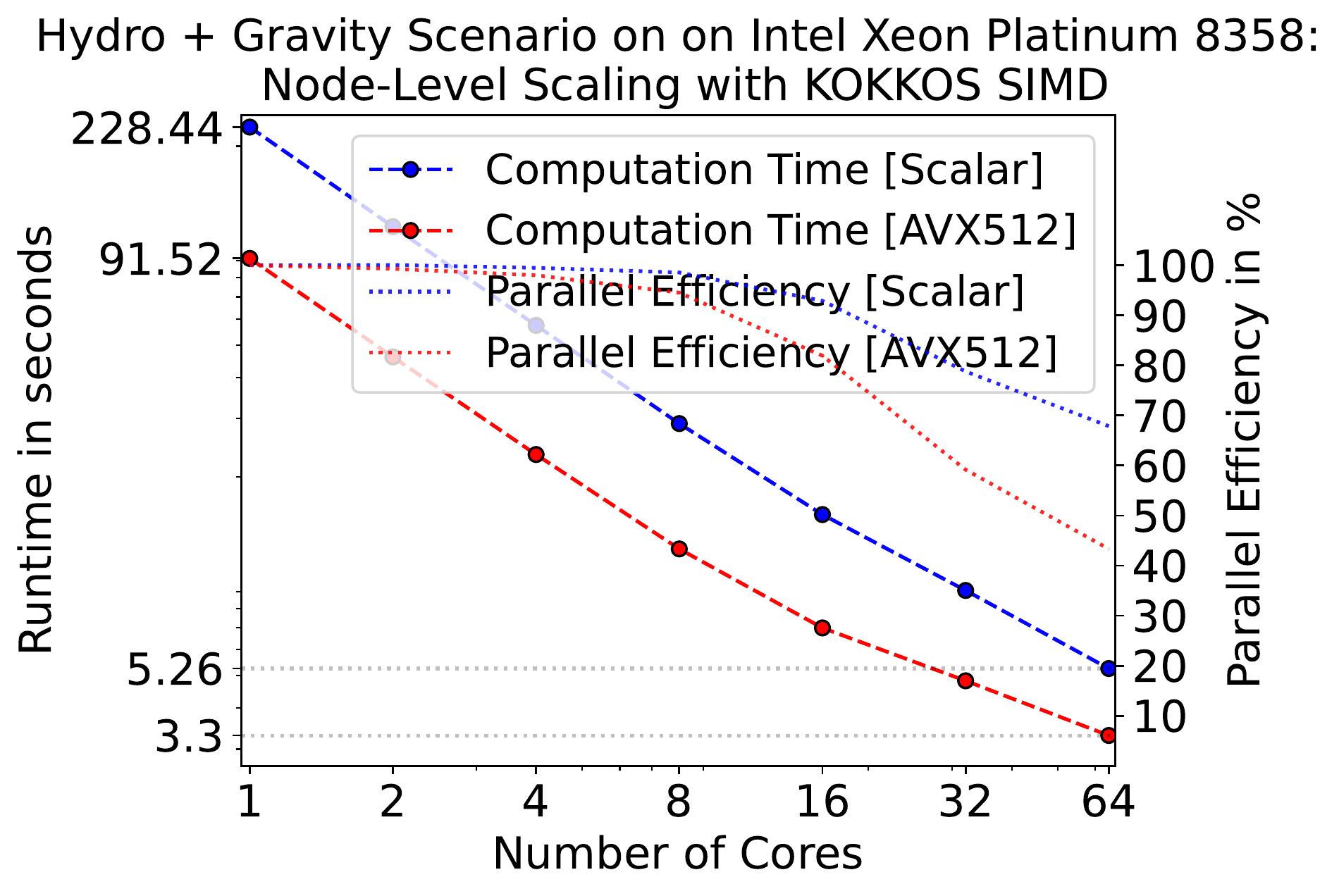} 
}
\hspace*{-0.3cm} 
  \subfloat[\label{fig:fixed_sve}On A64FX (using SVE)] {
\centering
  \includegraphics[width=.242\textwidth]{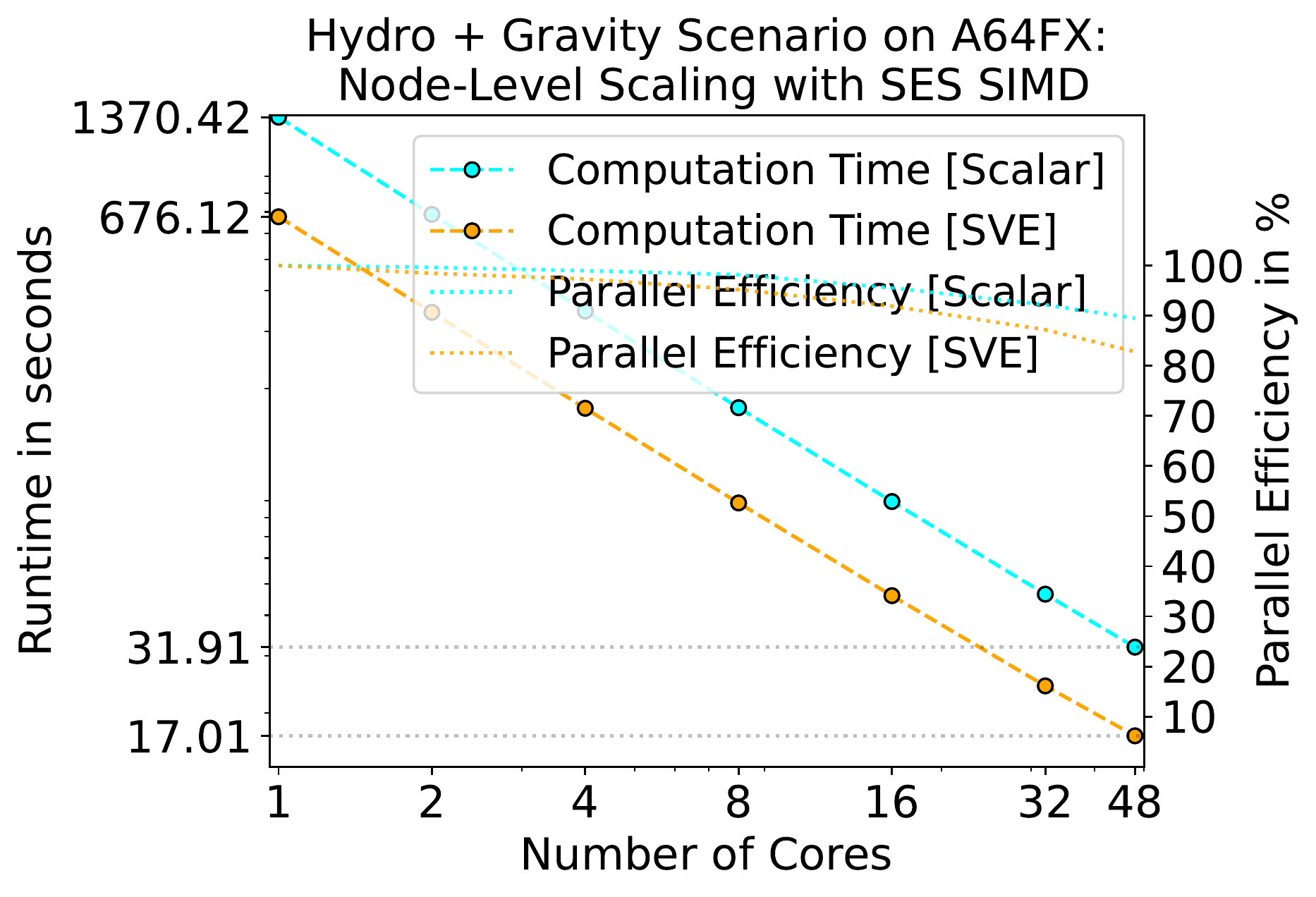} 
}  
\caption{Improving the overall A64FX and Intel runtimes shown in Figure~\ref{fig:node-level-scaling} by using more tasks per multipole kernel launch.}
\label{fig:fixed}
\end{figure}

Given that our compute kernels are working on one sub-grid each, we usually just launch one HPX task per Kokkos kernel, as this has been shown to work best with Octo-Tiger in previous work~\cite{daiss2021beyond}.
However, these past tests were done on CPUs with lower core counts.
On the CPUs used in this work with larger core counts, we can improve the performance by using more tasks, thus having more, albeit smaller work packages available during the tree traversal, which helps to keep all cores busy.
As we use the HPX execution space within Kokkos, this is a simple change in the kernel launch where we have to specify a different tiling configuration. Hence, we can easily split the multipole kernel (the most compute-intensive compute kernel per sub-grid) into more tasks and see how the performance is impacted by this.
By splitting this multipole kernel into more than one task per sub-grid using the Kokkos HPX execution space (into 16 tasks to be precise), we can achieve a better runtime, as can be seen in Figure~\ref{fig:fixed}. Here, we can reduce the overall computation runtime of the scenario from $20.67$s (see Fig.~\ref{fig:node-level-scaling-sve}) to $17.01$s (see Fig.~\ref{fig:fixed_sve}) on the A64FX ARM node just with this multipole change. This also helped on the Intel CPU, reducing the runtime to $3.3$s (Fig.~\ref{fig:fixed_intel}). However, this yielded no noticeable benefit on the AMD EYPC machine, as the bottleneck here appears to be different.

Scalability on the AMD EPYC machine (Fig.~\ref{fig:node-level-scaling-epyc}) proved to be troublesome: Beyond $64$ cores, the runtime deteriorates.
At first, we suspected that given our small scenario, the CPU is simply starved, but this is not the case; we see a similar parallel efficiency when running a larger scenario. 
Looking again at the mean kernel runtimes themselves instead, we noticed that the mean runtime of some kernels stay the same when moving to $128$ cores (as it should be, since each kernel is executed by one core by default). However, the mean runtime for other kernels (notably the \texttt{flux} kernel) increases.
This might be caused by inter-socket memory accesses hitting a bandwidth limit; however, at the time of submission, the profiling effort here is still ongoing and subject to future work.
%For now, the EPYC\textsuperscript{\texttrademark} shows the current limit of node-level scaling with Octo-Tiger.
For now, the EPYC shows Octo-Tiger's current node-level scaling limit. 

To summarize, using our SIMD changes, the node-level scaling looks acceptable on the Intel and good on the ARM node. Changing the launch configuration of the \texttt{Multipole} Kokkos kernel to use more tasks yielded additional benefits.
%The results caused us to further investigate the node-level scaling on the EPYC\textsuperscript{\texttrademark} node, though (as the core count per machine will only increase in future machines)
%We plan to investigate and improve the performance on A64FX to better match that of the Intel node. 
%We further plan to look into the bottleneck hindering the  node-level scaling on the EPYC\textsuperscript{\texttrademark} node.

\subsection{Test 3: Hydro-only Speedups}
As a last test, we briefly investigate the performance with a hydro-only scenario, as the most recent change to the compute kernels was the introduction of the SIMD types to the \texttt{flux} and \texttt{reconstruct} kernels.
This ensures that the only SIMD speedup we will see in this test will be due to our new implementation of those kernels, and not due to some existing SIMD implementation in the gravity solver.

We use the Sedov Blast Wave scenario, another one of Octo-Tiger's standard tests, as this scenario also has an analytical solution.
%We use the Sedov Blast Wave scenario\footnote{Octo-Tiger parameters: \texttt{--config\_file=blast.ini} \texttt{--unigrid=1} \texttt{--max\_level=3} \texttt{--stop\_step=10} \texttt{--correct\_am\_hydro=0} \texttt{--disable\_output=1} }, another one of standard Octo-Tiger's tests, as this scenario also has an analytical solution.
As we turned off the AMR, this scenario again contains $512$ leaf sub-grids, leading to the same overall grid as we had for the previous "rotating star" tests. This ensures that the computation times shown here should be relatable to the time the hydro solver part required in the previous two tests.
The only parts with explicit SIMD vectorization in this hydro test are the \texttt{reconstruct} and \texttt{flux} kernels, the rest of the solver and surrounding application do not benefit from switching to the SIMD types. 
Consequently, the computation time speedups themselves are lower than the speedups of the individual kernels as seen previously in Figure~\ref{fig:simd-speedup}.
We further include the old non-Kokkos hydro implementation here, to make sure that there are no performance regressions. This previous hydro implementation was only relying on HPX for multicore usage, but was incompatible with GPUs, hence the conversion to Kokkos in the first place.

\begin{figure}[h]
\centering
\subfloat[\label{fig:hydro_one_cores}Using one core] {
  \centering
  \includegraphics[width=.237\textwidth]{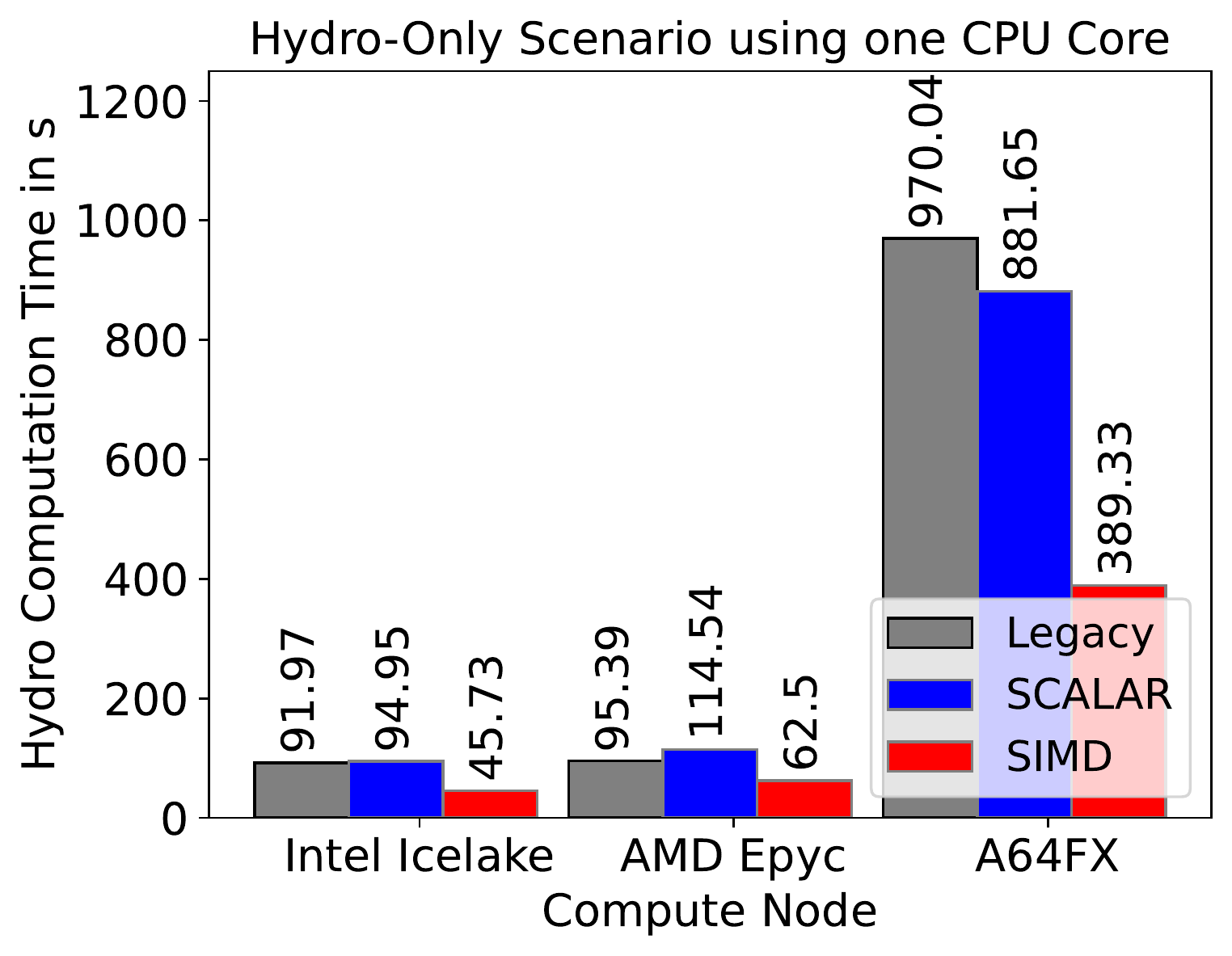}  
  %\caption{Using one core}
}
  \subfloat[\label{fig:hydro_all_cores}Using multiple cores] {
  \centering
  \includegraphics[width=.237\textwidth]{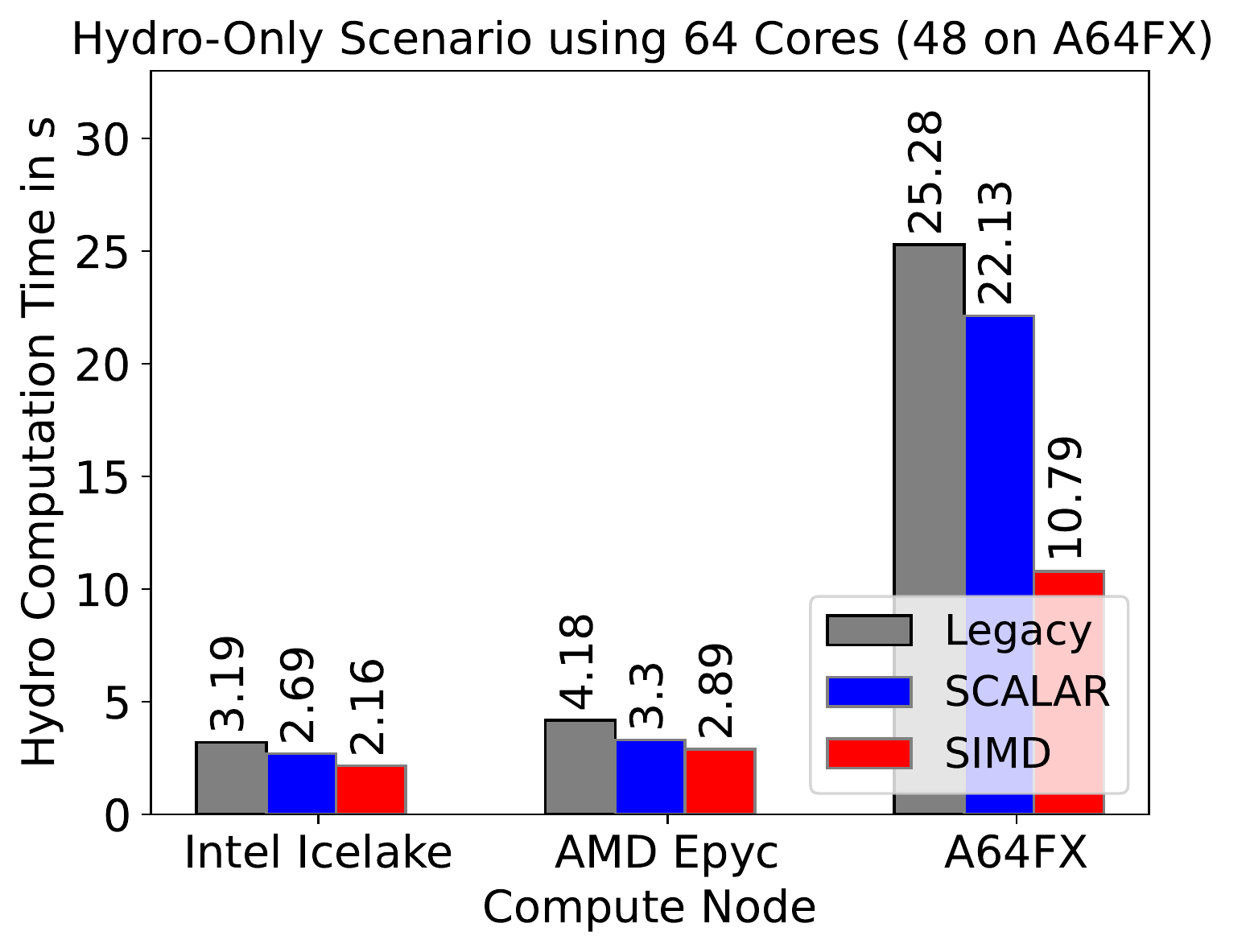}  
  %\caption{Using 64 cores each}
}
\caption{Running the Hydro-Only Sedov Blast Wave scenario using legacy hydro implementation, and the Kokkos hydro implementation using either scalar types or SIMD types (AVX512, AVX and SVE respectively). Left is the computation runtime on only one core, on the right is the same scenario when scaled to 64 cores (48 on A64FX).}
\label{fig:hydro}
\end{figure}

The results of this test can be found in Figure~\ref{fig:hydro}.
As expected, the best runs were those with SIMD types. There is no significant penalty when switching from the legacy kernels to Kokkos, even when just using the scalar types.  
Using SIMD the best overall speedup over the old implementation is reached on the A64FX machine.
On the Intel Icelake CPU, we took a look at the APEX data to investigate the speedup. 
Here, we spend a lot of time in the handling of the ghost cells and comparatively lesser percentage of the runtime in the compute kernels, resulting in the lower overall SIMD speedup shown here.

\section{Related Work}
\label{sec:related:work}

\subsection{Asynchronous many-task systems (AMT)}
HPX uses light-weight user threads to expose concurrency on top of system threads. Other notable solutions with this approach are: Chapel~\cite{chamberlain2007parallel}, Charm\texttt{++}~\cite{kale1993charm++}, PaRSEC~\cite{bosilca2013parsec}, Uintah~\cite{germain2000uintah} and Legion~\cite{bauer2012legion}. For a more detailed comparison of AMT's, like Cilk Plus, OpenMP, Intel TBB, Qthreads, StarPU, GASPI, Chapel, Charm\texttt{++} and HPX, we refer to~\cite{thoman2018taxonomy}.

\subsection{C\texttt{++} Performance Portability}
There are several options for performance portability using C\texttt{++}. First, there is Raja~\cite{beckingsale2019raja} providing CUDA, HIP, OpenMP and TBB backends. Second, there is SYCL providing hipSYCL, neoSYCL, triSYCL and TBB.
Finally, there is Thrust providing CUDA, TBB and OpenMP. For a comparison between Kokkos and SYCL, we refer to~\cite{hammond2019comparative}. 
However, none other than Kokkos provides an HPX backend.

\subsection{SIMD}
To achieve vectorization, one can rely on the auto-vectorization provided by the compiler~\cite{naishlos2004autovectorization,finkel2012autovectorization}. However, this is more like a black box, and the programmer does not have full control. More user control is provided by explicit SIMD, e.g.\ UME::SIMD~\cite{karpinski2017high}, Vc~\cite{kretz2012vc} or \lstinline{std::simd}. Here, the programmer uses types, like \lstinline{std::exerimental::simd}, which represent vector registers. 
On the other hand, there is implicit SIMD, where the programmer adds directives (such as pragmas) to indicate if code should get vectorized by the compiler~\cite{pohl2016evaluation}.
A well known example of this is are the SIMD pragmas in OpenMP.
%Here, during the build process, scalar types within compute kernels might get transformed to SIMD operation if there might be a performance benefit. 
%However, this is heuristic. 
We use explicit SIMD to have the most control. 

\subsection{SVE}
SVE is the new vectorization standard~\cite{armsve} introduced in ARM v8 in the Fugaku Fujitsu Supercomputer. This follows the VLA (vector length agnostic)  programming model, where the vector length is set dynamically from $128$ bits to $2048$ bits in increments of $128$ bits. 
SVE vectorization can be utilized in multiple ways as well, by using compiler auto-vectorization or by using explicit vectorization using fixed sized SIMD types. 
%SVE Vectorization can also be used with a fixed sized (length), which is useful for implementing explicit vectorization.
For more details, we refer to~\cite{7924233}.
% \todo[inline]{This is not really citing any related work - should we keept it / remove it / or move it to a different section?}

\section{Conclusion}
\label{sec:conclusion}
The presented execution model, a combination of HPX, Kokkos and explicit vectorization, proves to be versatile in its use in Octo-Tiger.
We can now easily switch both the SIMD library (Kokkos SIMD or \lstinline{std::experimental::simd}) and the used SIMD extensions.
Moreover, adjusting the number of HPX tasks that a Kokkos kernels gets split into further helps in coping with larger CPU core counts during the FMM tree traversals.

We tested this execution model (and our changes) on three CPU architectures using Octo-Tiger as a real-world benchmark.
Overall, Octo-Tiger benefits from the changes presented here:
Introducing the new SIMD implementation for the two hydro kernels, we experienced a speedup on all platforms. 
On A64FX, our target platform for the changes presented here, we measured the largest improvements. 
Here, switching the SIMD types to our new SVE types and increasing the number of HPX tasks clearly improved the performance. 
Both of these options can be simply changed during compile time.
This allows us to run more efficiently on A64FX using SIMD instead of the scalar types. 
However, taking a closer look at the node-level scaling, we unfortunately uncovered a pre-existing bottleneck on the AMD EPYC platform when going from 64 to 128 cores.

It is noteworthy that the presented execution approach is application-independent. 
The added SVE SIMD types in particular can be useful for other applications and will benefit the community beyond Octo-Tiger. 
%As for Octo-Tiger itself: Due to the integration of, \lstinline{std::experimental::simd} it will be easier in the future to switch SIMD backends, which are added to the \lstinline{std::experimental::simd} framework and the SIMD speedups obtained by the explicit vectorization are of course already useful on their own.

Taking a look at Octo-Tiger's current development snapshot in this work, for our next steps, we plan to examine the node-level scalability limit observed on the AMD EPYC node, and to improve the multitude of compute kernels in Octo-Tiger with SVE to achieve better SIMD speedups for each of them.
%Taking a look at Octo-Tiger's current development snapshot in this work, for our future work we plan to take a closer look at the lackluster parallel efficiency on the AMD EPYC node and further improve the compute kernels within Octo-Tiger to achieve better invidual SIMD speedups, especially using SVE.
Afterwards, we plan to run Octo-Tiger on NERSC's Permutter, Riken's Fugaku, and CSC's LUMI to gather more performance measurements, both node-level and distributed, on a diverse set of machines.

\section{Acknowledgment}
{\footnotesize
The authors would like to thank Stony Brook Research Computing and Cyberinfrastructure, and the Institute for Advanced Computational Science at Stony Brook University for access to the innovative high-performance Ookami computing system, which was made possible by a \$5M National Science Foundation grant (\#1927880).}

\section*{Copyright notice}
\textcopyright 2022 IEEE. Personal use of this material is permitted.
  Permission from IEEE must be obtained for all other uses, in any current or future 
  media, including reprinting/republishing this material for advertising or promotional 
  purposes, creating new collective works, for resale or redistribution to servers or 
  lists, or reuse of any copyrighted component of this work in other works.

\bibliographystyle{IEEEtran}
\bibliography{references.bib}

%\vspace{12pt}

% LaTeX template for the Supercomputing Conference series Artifact Description (AD) appendix  
% V20180327
% (C)opyright 2018

% Derived with permission by Michael Heroux (Sandia National Laboratories, St. John's University, MN)
% from ae-20160509.tex 
% written by Grigori Fursin (cTuning foundation, France and dividiti, UK) 
% and Bruce Childers (University of Pittsburgh, USA)
% (C)opyright 2014-2016

% acmart is available at https://www.acm.org/publications/proceedings-template
%\documentclass[sigconf,twocolumn]{acmart}
% IEEETrans is available at https://www.ieee.org/conferences_events/conferences/publishing/templates.html

%% removed as per package comment before \ appendices
%\documentclass{IEEEtran}
%
%\usepackage[hidelinks]{hyperref}
%\usepackage{listings}
%\begin{document}
%
%\special{papersize=8.5in,11in}
%
%%%%%%%%%%%%%%%%%%%%%%%%%%%%%%%%%%%%%%%%%%%%%%%%%%%%
% When adding this appendix to your paper, 
% please remove above part
%%%%%%%%%%%%%%%%%%%%%%%%%%%%%%%%%%%%%%%%%%%%%%%%%%%%

\appendices

\section{Artifact Description Appendix: From Merging Frameworks to Merging Stars: Experiences using HPX, Kokkos and SIMD Types}

%%%%%%%%%%%%%%%%%%%%%%%%%%%%%%%%%%%%%%%%%%%%%%%%%%%%%%%%%%%%%%%%%%%%%
\subsection{Abstract}
This description contains the instructions needed to run the experiments described in the ESPM2 2022 Paper "From Merging Frameworks to Merging Stars: Experiences using HPX, Kokkos and SIMD Types".
Here, we briefly describe the utilized real-world application (Octo-Tiger), its required dependencies and the respective versions that we used to obtain the results in the paper.
We further describe the utilized hardware and Octo-Tiger scenarios.
Additionally, we provide details regarding how to build Octo-Tiger (and all required dependencies), where to find the required experiment scripts and how to run them to recreate our results (and plots).
%%%%%%%%%%%%%%%%%%%%%%%%%%%%%%%%%%%%%%%%%%%%%%%%%%%%%%%%%%%%%%%%%%%%%
\subsection{Description}

\subsubsection{Check-list (artifact meta information)}

%{\em Fill in whatever is applicable with some informal keywords and remove the rest}

{\small
\begin{itemize}
  \item {\bf Algorithm: Octo-Tiger primarily uses the Fast-Multipole Method for the gravity solver and finite volumes for the hydro solver}
  \item {\bf Program: Octo-Tiger}
  \item {\bf Compilation: g++11.2 (buildscripts available)}
  \item {\bf Hardware: Intel Xeon CPU, AMD EYPC CPU, A64FX CPU}
  \item {\bf Output: Python script to recreate the plots is available}
  \item {\bf Experiment workflow: Iterate over cores and SIMD extensions (Bash scripts to re-run these experiments are available)}
  \item {\bf Experiment customization: Modified Kokkos and Octo-Tiger (commits and patches are available)}
  \item {\bf Publicly available?: Yes (GitHub) }
\end{itemize}
}

\subsubsection{How software can be obtained (if available)}
Octo-Tiger can be obtained on GitHub:

\noindent \url{https://github.com/STEllAR-GROUP/octotiger}

\noindent We used two Octo-Tiger commits:
\begin{itemize}
  \item \textit{2461163996c576a5d43b4cd1a7f8c295d446f925} (for using more tasks in the multipole kernels to generate Figure~\ref{fig:fixed})
\item \textit{9437e172264a4c7b3f24ec4ccd4c380feba26755}  (for all other experiments)
\end{itemize}

%{\em Obligatory if the paper contains computational results.}

\subsubsection{Hardware dependencies}
We run our experiments (tests) on three different machines (nodes):
\begin{enumerate}
  \item One Fujitsu A64FX\textsuperscript{\texttrademark} CPU @ 1.80 GHz on Ookami (48 cores).
\item One university server (at LSU) containing a two-socket Intel\textsuperscript{\textregistered}~Xeon\textsuperscript{\textregistered} Platinum 8358 CPU @ 2.60GHz (64 cores)
\item One university server (at LSU) containing a two-socket AMD\textsuperscript{\textregistered} EPYC\textsuperscript{\texttrademark} 7H12 CPU @ 2.60GHz (128 cores) 
\end{enumerate}

\noindent Memory requirements for the scenarios we use are minimal and should not exceed 4 GB for each scenario itself.

\subsubsection{Software dependencies}
Octo-Tiger has multiple, mandatory software dependencies:
\begin{itemize}
\item HPX: 1.8.0
\item Kokkos (on the Intel and AMD nodes): 

\textit{2640cf70de338618a7e4fe10590b06bc1c239f4c} 

\item Kokkos (on Ookami): 

\textit{596bb0b1b} 

\item HPX-Kokkos:

\textit{20a44967c742f5a7670b4dff9658d9973bf849f2} 

URL: \url{https://github.com/STEllAR-GROUP/hpx-kokkos}
\item CPPuddle (on the Intel and AMD nodes): 

\textit{6127562897dc2940869d744d490d0eb7b6fa37bc} 

\item CPPuddle (on Ookami): 

\textit{f1eed375685981b59723e5592961c5c774789a20} 

URL: \url{https://github.com/SC-SGS/CPPuddle}
\item Vc: 1.4.1
\item Jemalloc: 5.2.1
\item Silo: 4.10.2
\item HDF5: 1.8.12
\item Boost: 1.75 (on the Intel and AMD nodes)
\item Boost: 1.78 (on Ookami)
\item SVE types library: 

\textit{e3b2fb8d7bfda5d6eda90efee29fafdcbe895a25} 

URL: \url{https://github.com/srinivasyadav18/sve}
\end{itemize}
All dependencies are available online! 
We have added links for the lesser known ones though, to ease reproducibility.
We use slightly different versions (as outlined in the list above) on Ookami to get the toolchain compiled on this different system.
For ease of use, there are buildscripts available at this URL:

\noindent \url{https://github.com/STEllAR-GROUP/OctoTigerBuildChain}

We use two additional patchfiles, one for Kokkos and one for HPX-Kokkos.
Kokkos has been customized based on the commits specified here, to allow the single-task optimization mentioned in Section~\ref{sec:changes:kernel}).
This has not yet been upstreamed. However, the buildscripts contain the patchfile for this (kokkos-single-task.patch) and apply the patch automatically during building.
Accordingly, HPX-Kokkos requires a patch adapting to this which is also included in the buildscripts (sync.patch) and applied automatically for building.
These changes will be upstreamed eventually, but for the given commits above these two patches are required.

%The version differences on Ookami are changes to get gcc/11.2 to compile the toolchain there with C++20 as this is required for the SVE types.

\subsubsection{Datasets}
No additional datasets need to be downloaded! 
The inputfile required for the "rotating star" scenario can be created using the binary \texttt{gen\_rotating\_star\_init} which will be build automatically when building Octo-Tiger (to be found inside the tools subdirectory within the Octo-Tiger build).
Further configuration files (blast.ini and rotating\_star.ini) are part of the Octo-Tiger source repository (in the test\_problems subdirectory) and supply the exact configuration for each scenario.

Using these configuration files, the rotating star scenario can be started with the following runtime parameters (adapt the paths to the ini and binary files if required):
\begin{lstlisting}[language=bash, frame=single, basicstyle=\ttfamily\footnotesize]
$ ./build/octotiger/build/octotiger \
--config_file=rotating_star.ini \
--unigrid=1 --max_level=3 --stop_step=10 \
--correct_am_hydro=0 --theta=0.34 --disable_output=1
\end{lstlisting}

\noindent Respectively, for the Sedov Blast Wave scenario:
\begin{lstlisting}[language=bash, frame=single, basicstyle=\ttfamily\footnotesize]
$ ./build/octotiger/build/octotiger \
--config_file=blast.ini \
--unigrid=1 --max_level=3 --stop_step=10 \
--correct_am_hydro=0 --disable_output=1
\end{lstlisting}

\noindent For each scenario, we additionally configure whether the KOKKOS or LEGACY kernels are used with the following parameters:
\begin{lstlisting}[language=bash, frame=single, basicstyle=\ttfamily\footnotesize]
--hydro_host_kernel_type=KOKKOS
--multipole_host_kernel_type=KOKKOS
--monopole_host_kernel_type=KOKKOS
\end{lstlisting}
Lastly, the number of utilized worker threads (and thus CPU cores) can be steered with the parameter \lstinline[language=bash, basicstyle=\normalsize]{--hpx:threads}. 

%%%%%%%%%%%%%%%%%%%%%%%%%%%%%%%%%%%%%%%%%%%%%%%%%%%%%%%%%%%%%%%%%%%%%
\subsection{Installation}
There are buildscripts available to ease building Octo-Tiger with the software dependencies given earlier:

\noindent \url{https://github.com/STEllAR-GROUP/OctoTigerBuildChain}
These scripts will automatically download and build the required dependencies, as well as Octo-Tiger itself.

\subsubsection{Configuring on the Intel Node}
\begin{itemize}
    \item To install on the LSU Intel machine, use the buildscripts commit: \textit{ca909534c3c93f7fddb77f48d15c42dba558f0e0} (or the branch espm2\_icelake\_build)
    \item Make sure the modules gcc/11.2.1 and hwloc 2.4.1 are loaded!
\end{itemize}

\subsubsection{Configuring on the AMD Node}
\begin{itemize}
\item To install on LSU AMD machine, use the buildscripts commit: \textit{a53f1413105dd543560e30c77c710d9f32513fa0} (or the branch espm\_epyc)
\item Make sure the modules gcc/11.2.1 and hwloc 2.4.1 are loaded!
\end{itemize}
\subsubsection{Configuring on the Ookami A64FX Node}
\begin{itemize}
    \item To install on Ookami, use the buildscripts commit: \textit{6c9f6361c0942cb5ceb2dde986c42614cc1102c0} (or the branch espm2\_ookami)
    \item Make sure the modules gcc/11.2.0 cmake/3.22.1 and hwloc/2.4.1 are loaded!
\end{itemize}

\subsubsection{Building (on all machines)} 
\begin{itemize}
\item First, load the modules and configure the buildscripts as outlined in the last three subsections (depending on your machine).
\item Afterwards, run the following command within the root directory of the buildscripts:
\end{itemize}
\begin{lstlisting}[language=bash, frame=single, basicstyle=\ttfamily\footnotesize]
$ ./build-all.sh Release with-CC \
without-cuda without-mpi without-papi \
with-apex with-kokkos with-simd \
with-hpx-backend-multipole \
with-hpx-backend-monopole \
with-hpx-cuda-polling without-otf2 \
boost hdf5 silo jemalloc vc hpx \
kokkos cppuddle octotiger
\end{lstlisting}

\subsubsection{Changing the SIMD types}
\begin{itemize}
\item To switch SIMD types and libraries manually, the following Octo-Tiger CMAKE flags can be used (modify them in \texttt{build-octotiger.sh} in the buildscripts):
\begin{itemize}
    \item OCTOTIGER\_KOKKOS\_SIMD\_LIBRARY
    \item OCTOTIGER\_KOKKOS\_SIMD\_EXTENSION
\end{itemize}
\item Note: The experiment bash scripts will do this automatically, so there should be no need to do this manually!
\end{itemize}

%%%%%%%%%%%%%%%%%%%%%%%%%%%%%%%%%%%%%%%%%%%%%%%%%%%%%%%%%%%%%%%%%%%%%
\subsection{Experiment workflow}

The bash scripts to perform the experiments on the Intel and AMD node can be found on GitHub:
\noindent \url{https://github.com/G-071/octotiger-performance-tests/tree/master/rostam/node-level-scaling-tests}

The experiment \texttt{test\_cpu\_performance.sh} gathers all the runtime data when being launched (after building Octo-Tiger) within the root directory of the buildscripts.
This experiment iterates over all core and SIMD configurations, generating a (csv) log file containing the runtime data for the application and the compute kernels.
There are multiple scenario files available within the GitHub repository.
These scenario files mostly contain the core configuration and specify the SIMD types to be used on each machine.
To make the experiments work, one needs to pass one of these scenario files (the rotating star one or the blast wave one for our experiments) to the bash script as the first parameter. 
%The experiment script \texttt{test\_cpu\_performance.sh} needs to be run from the root directory of the buildscripts.
Afterwards, run the second experiment script \texttt{test\_hydro\_speedup.sh} (also within the buildscripts root directory) to obtain the hydro-only runtime data required for the last figure in the paper.

Due to a change in the output format (requiring adaptations to the bash scripts), the respective scripts for Ookami can also be found on GitHub but in a different directory:
\noindent \url{https://github.com/G-071/octotiger-performance-tests/tree/master/ookami/node-level-scaling-tests}

\noindent However, the process of running the tests is the same as for the Intel and AMD nodes otherwise.

%%%%%%%%%%%%%%%%%%%%%%%%%%%%%%%%%%%%%%%%%%%%%%%%%%%%%%%%%%%%%%%%%%%%%
\subsection{Evaluation and expected result}

To evaluate the data and plot the results, there is a python script (\texttt{plot\_node\_level\_scaling.py}) available at:

\noindent \url{https://github.com/G-071/octotiger-performance-tests/tree/master/rostam/node-level-scaling-tests/plot} 

This python script can be used to generate the graphs in the paper, using the runtime data obtained in the last subsection!
The script requires pandas, numpy and matplotlib to be installed.
To run it, one needs to pass the file containing the csv results from a machine via the parameter \texttt{--filename} (pass the output file generated by \texttt{test\_cpu\_performance.sh}).
Additionally, pass the utilized SIMD extension via \texttt{--simd\_key} (for example AVX512).
This should plot both the node-level scaling graphs (Figure~\ref{fig:node-level-scaling}) and the SIMD speedup graphs (Figure~\ref{fig:simd-speedup}) and immediately store them as PDFs in the same directory.
To plot the last figure (Figure~\ref{fig:hydro}), the script further expects the hydro-only results (obtained by running \texttt{test\_hydro\_speedup.sh}) to be present in the following three files:
\texttt{icelake\_legacy\_test.data},
\texttt{epyc\_legacy\_test.data}, 
\texttt{arm\_legacy\_test.data}

\noindent The expected results are plots similar to what is shown in the paper!

%%%%%%%%%%%%%%%%%%%%%%%%%%%%%%%%%%%%%%%%%%%%%%%%%%%%%%%%%%%%%%%%%%%%%
\subsection{Experiment customization}
Kokkos gets modified with a patch as outlined earlier.
To generate Fig.~\ref{fig:fixed}, a different Octo-Tiger commit (as mentioned) has to be used before running the experiment scripts!

%%%%%%%%%%%%%%%%%%%%%%%%%%%%%%%%%%%%%%%%%%%%%%%%%%%%%%%%%%%%%%%%%%%%%
%\subsection{Notes}

%%%%%%%%%%%%%%%%%%%%%%%%%%%%%%%%%%%%%%%%%%%%%%%%%%%%
% When adding this appendix to your paper, 
% please remove below part
%%%%%%%%%%%%%%%%%%%%%%%%%%%%%%%%%%%%%%%%%%%%%%%%%%%%

%\end{document}

\end{document}

\special{papersize=8.5in,11in}

%%%%%%%%%%%%%%%%%%%%%%%%%%%%%%%%%%%%%%%%%%%%%%%%%%%%
% When adding this appendix to your paper, 
% please remove above part
%%%%%%%%%%%%%%%%%%%%%%%%%%%%%%%%%%%%%%%%%%%%%%%%%%%%

\appendices

\section{Artifact Description Appendix: From Merging Frameworks to Merging Stars: Experiences using HPX, Kokkos and SIMD Types}

%%%%%%%%%%%%%%%%%%%%%%%%%%%%%%%%%%%%%%%%%%%%%%%%%%%%%%%%%%%%%%%%%%%%%
\subsection{Abstract}
This description contains the instructions needed to run the experiments described in ESPM2 2022 Paper "From Merging Frameworks to Merging Stars: Experiences using HPX, Kokkos and SIMD Types".
It briefly describes the utilized real-world application (Octo-Tiger), its required dependencies and the respective versions that were used for obtaining the results in the paper. Furthermore, it contains descriptions of the utilized hardware and Octo-Tiger scenarios.
It concludes with instructions on how to build Octo-Tiger (with the required dependencies) and where to find the experiment (and plot) scripts to recreate said results.
%Octo-Tiger, a large-scale 3D AMR code for the merger of stars, uses a combination of HPX, Kokkos and explicit SIMD types, aiming to achieve performance-portability for a broad range of heterogeneous hardware.
%However, on A64FX CPUs, we encountered several missing pieces, hindering performance by causing problems with the SIMD vectorization.
%Therefore, we add std::experimental::simd as an option to use in Octo-Tiger's Kokkos kernels alongside Kokkos SIMD, and further add a new SVE (Scalable Vector Extensions) SIMD backend.
%Additionally, we amend missing SIMD implementations in the Kokkos kernels within Octo-Tiger's hydro solver.
%We test our changes by running Octo-Tiger on three different CPUs: An A64FX, an Intel Icelake and an AMD EPYC CPU, evaluating SIMD speedup and node-level performance.
%We get a good SIMD speedup on the A64FX CPU, as well as noticeable speedups on the other two CPU platforms. However, we also experience a scaling issue on the EPYC CPU.

%%%%%%%%%%%%%%%%%%%%%%%%%%%%%%%%%%%%%%%%%%%%%%%%%%%%%%%%%%%%%%%%%%%%%
\subsection{Description}

\subsubsection{Check-list (artifact meta information)}

%{\em Fill in whatever is applicable with some informal keywords and remove the rest}

{\small
\begin{itemize}
  \item {\bf Algorithm: Octo-Tiger primarily uses the Fast-Multipole Method for the gravity solver and finite volumes for the hydro solver}
  \item {\bf Program: Octo-Tiger}
  \item {\bf Compilation: g++11.2}
  \item {\bf Hardware: Intel CPU, AMD CPU, A64Fx CPU}
  \item {\bf Output: Python Script to re-run the plots is available}
  \item {\bf Experiment workflow: Iterate over cores and SIMD extensions, Bash scripts to re-run experimentation are available}
  \item {\bf Experiment customization: Modified Kokkos and Oco-Tiger, commits and patches are available}
  \item {\bf Publicly available?: GitHub }
\end{itemize}
}

\subsubsection{How software can be obtained (if available)}
Octo-Tiger can be obtained on GitHub:
\url{https://github.com/STEllAR-GROUP/octotiger}

We used two Octo-Tiger commits:
\begin{itemize}
\item \textit{2461163996c576a5d43b4cd1a7f8c295d446f925} (for using more tasks in the multipole kernels to generate Figure 5.)
\item \textit{9437e172264a4c7b3f24ec4ccd4c380feba26755}  (for all other tests)
\end{itemize}

%{\em Obligatory if the paper contains computational results.}

\subsubsection{Hardware dependencies}
We run our tests on three different machines:
\begin{enumerate}
\item One A64Fx node on Ookami.
\item One university server (at LSU) containing a two-socket Intel® Xeon® Platinum 8358 Processor CPU (64 cores)
\item One university server (at LSU) containing a two-socket AMD Epyc 7H12 (128 cores) 
\end{enumerate}

Memory requirements for the scenarios we use are minimal and should not exceed 4 GB for each scenario itself.

\subsubsection{Software dependencies}
Octo-Tiger has multiple, mandatory software dependencies:
\begin{itemize}
\item HPX: 1.8.0
\item Kokkos: 

\textit{2640cf70de338618a7e4fe10590b06bc1c239f4c} 

(on Intel and AMD)

\textit{596bb0b1b} 

(on Ookami)
\item HPX-Kokkos:

\textit{20a44967c742f5a7670b4dff9658d9973bf849f2} 

URL: \url{https://github.com/STEllAR-GROUP/hpx-kokkos}
\item CPPuddle: 

\textit{6127562897dc2940869d744d490d0eb7b6fa37bc} 

(on Intel and AMD)

\textit{f1eed375685981b59723e5592961c5c774789a20} 

(on Ookami)

URL: \url{https://github.com/SC-SGS/CPPuddle}
\item Vc: 1.4.1
\item Jemalloc: 5.2.1
\item silo: 4.10.2
\item HDF5: 1.8.12
\item Boost: 1.75 (on Intel and AMD machine), 1.78 (on Ookami)
\item SVE type library: 

\textit{e3b2fb8d7bfda5d6eda90efee29fafdcbe895a25} 

URL: \url{https://github.com/srinivasyadav18/sve}
\end{itemize}
All dependencies are available online! We have added links for the lesser known ones though, to ease reproducibility.
For ease of use there are buildscripts available at this URL:

\url{https://github.com/STEllAR-GROUP/OctoTigerBuildChain}
Kokkos has been customized based on the commits specified here, to allow the single-task optimization mentioned in section V A).
This has not yet been upstreamed. However, the buildscripts contain the patchfile for this (kokkos-single-task.patch) and apply the patch automatically during building.
Accordingly, HPX-Kokkos requires a patch adapting to this which is also included in the buildscripts (sync.patch) and applied automatically for building.
These changes will be upstreamed eventually, but for the given commits here the patches are required.

We use slightly different versions (as outlined in the list above) on Ookami to get the toolchain compiled on this different system.

%The version differences on Ookami are changes to get gcc/11.2 to compile the toolchain there with C++20 as this is required for the SVE types.

\subsubsection{Datasets}
No additional datasets are required! 
The inputfile required for the "rotating star" scenario can be created using the binary \texttt{gen\_rotating\_star\_init} which will be built automatically when building Octo-Tiger (to be found inside the tools subdirectory within the Octo-Tiger build).
The configuration files blast.ini and rotating\_star.ini are also part of the Octo-Tiger repository (in the test\_problems sub directory).

Using these configuration files, the Rotating Star scenario can be started with these parameters (adapt path to the ini files):

Parameters: \texttt{
--config\_file=rotating\_star.ini} \texttt{--unigrid=1} \texttt{--max\_level=3} \texttt{--stop\_step=10} \texttt{--correct\_am\_hydro=0} \texttt{--theta=0.34} \texttt{--disable\_output=1}
For the Sedov Blast Wave scenario:

Parameters: 
\texttt{--config\_file=blast.ini} \texttt{--unigrid=1} \texttt{--max\_level=3} \texttt{--stop\_step=10} \texttt{--correct\_am\_hydro=0} \texttt{--disable\_output=1}

Additional parameters steer whether the KOKKOS or LEGACY kernels are used:
\begin{itemize}
    \item \texttt{--hydro\_host\_kernel\_type=KOKKOS}
    \item \texttt{--multipole\_host\_kernel\_type=KOKKOS}
    \item \texttt{--monopole\_host\_kernel\_type=KOKKOS}
\end{itemize}

%%%%%%%%%%%%%%%%%%%%%%%%%%%%%%%%%%%%%%%%%%%%%%%%%%%%%%%%%%%%%%%%%%%%%
\subsection{Installation}
There are buildscripts available to ease building Octo-Tiger with the given software dependencies above:

\url{https://github.com/STEllAR-GROUP/OctoTigerBuildChain}
These scripts will automatically download and build the required dependencies, as well as Octo-Tiger itself.

\subsubsection{Configuring on the Intel Node}
\begin{itemize}
    \item To install on the LSU Intel machine, use the buildscripts commit: \textit{f889e5abaa8d709ac4319e38f38ada75d63a7d73} (or the branch espm2\_icelake\_build)
    \item Make sure modules gcc/11.2.1 and hwloc 2.4.1 are loaded!
\end{itemize}

\subsubsection{Configuring on the AMD Node}
\begin{itemize}
\item To install on LSU AMD machine, use the buildscripts commit: \textit{747a3a5e1dc71dab640d4989367292cf00fba49f} (or the branch espm\_epyc)
\item Make sure modules gcc/11.2.1 and hwloc 2.4.1 are loaded!
\end{itemize}
\subsubsection{Configuring on the Ookami A64Fx Node}
\begin{itemize}
    \item To install on Ookami, use the buildscripts commit: \textit{6c9f6361c0942cb5ceb2dde986c42614cc1102c0} (or the branch espm2\_ookami)
    \item Make sure the module gcc/11.2.0 cmake/3.22.1 and hwloc/2.4.1 are loaded
\end{itemize}

\subsubsection{Building (on all machines)}
First, load the modules and configure the buildscripts as outlined in the last three subsections (depending on your machine).
Afterwards, run the following command within the root directory of the buildscripts:
\begin{lstlisting}[language=bash, frame=single]
$ ./build-all.sh Release with-CC 
without-cuda without-mpi without-papi 
with-apex with-kokkos with-simd 
with-hpx-backend-multipole 
with-hpx-backend-monopole
with-hpx-cuda-polling without-otf2 
boost hdf5 silo jemalloc vc hpx 
kokkos cppuddle octotiger
\end{lstlisting}

\subsubsection{Changing the SIMD types}
To switch SIMD types and libraries manually, the Octo-Tiger CMAKE flags can be used (modify \texttt{build-octotiger.sh} in the buildscripts):
\begin{itemize}
    \item OCTOTIGER\_KOKKOS\_SIMD\_LIBRARY
    \item OCTOTIGER\_KOKKOS\_SIMD\_EXTENSION
\end{itemize}
Note: The experiment bash scripts will do this automatically, so there should not be a need to do this manually!

%%%%%%%%%%%%%%%%%%%%%%%%%%%%%%%%%%%%%%%%%%%%%%%%%%%%%%%%%%%%%%%%%%%%%
\subsection{Experiment workflow}

The scripts to perform the experiments on the Intel and AMD node can be found on GitHub:
\url{https://github.com/G-071/octotiger-performance-tests/tree/master/rostam/node-level-scaling-tests}

The experiment \texttt{test\_cpu\_performance.sh} gathers all the runtime data when being launched after building within the root directory of the buildscripts.

The experiment itself iterates over all core and SIMD configurations, generating a log file containing the runtime data for the application and the compute kernels in csv format.

There are multiple scenario files within the GitHub repository available. To make the experiment work, you need to pass one (the rotating star one or the blast wave one for our experiments) to the script. The scenario files mostly contain the core configuration and specify the SIMD types to be used on each machine.

The bash script \texttt{test\_cpu\_performance.sh} for experiment needs to be run from the root directory of the buildscripts.
Afterwards run the second bash script \texttt{test\_hydro\_speedup.sh}, also within the buildscripts root directory, to obtain the runtime data for the last figure in the paper.

Due to a change in the output format, the tests for Ookami can be found also on GitHub but in a different directory:

\url{https://github.com/G-071/octotiger-performance-tests/tree/master/ookami/node-level-scaling-tests}

However, the process of running the tests is the same as for the Intel and AMD node otherwise.

%%%%%%%%%%%%%%%%%%%%%%%%%%%%%%%%%%%%%%%%%%%%%%%%%%%%%%%%%%%%%%%%%%%%%
\subsection{Evaluation and expected result}

To evaluate the data and plot the results, there is a python script (\texttt{plot\_node\_level\_scaling.py}) available at:

\url{https://github.com/G-071/octotiger-performance-tests/tree/master/rostam/node-level-scaling-tests/plot} 

This python script can be used to plot the data obtained on each machine to generate the graphs in the paper! One needs to pass the file containing the csv results from the machines via \texttt{--filename=} (see \texttt{test\_cpu\_performance.sh}) in addition to the utilized SIMD extension \texttt{--simd\_key}, for example AVX512. The python script expects pandas, numpy and matplotlib to be available.
This should plot the node-level scaling graphs and the SIMD speedup graphs and immediately store them as PDFs in the same directory.
To plot the last figure, the script also expects the other results, obtained by running \texttt{test\_hydro\_speedup.sh} test, to be present in the files:
\begin{itemize}
    \item \texttt{icelake\_legacy\_test.data}
    \item \texttt{epyc\_legacy\_test.data}
    \item \texttt{arm\_legacy\_test.data}
\end{itemize}

The expected results are plots similar to what is shown in the paper!

%%%%%%%%%%%%%%%%%%%%%%%%%%%%%%%%%%%%%%%%%%%%%%%%%%%%%%%%%%%%%%%%%%%%%
\subsection{Experiment customization}
Kokkos gets modified with a patch as outlined above.
To generate Figure 5, at different Octo-Tiger commit (see above) has to be used before running the experiment script!

%%%%%%%%%%%%%%%%%%%%%%%%%%%%%%%%%%%%%%%%%%%%%%%%%%%%%%%%%%%%%%%%%%%%%
%\subsection{Notes}

%%%%%%%%%%%%%%%%%%%%%%%%%%%%%%%%%%%%%%%%%%%%%%%%%%%%
% When adding this appendix to your paper, 
% please remove below part
%%%%%%%%%%%%%%%%%%%%%%%%%%%%%%%%%%%%%%%%%%%%%%%%%%%%